\pdfoutput=1
\documentclass[twoside,jaspp]{ametsoc}  

\usepackage[pdftex]{graphicx}   

\usepackage[latin9]{inputenc}
\usepackage[letterpaper]{geometry}
\usepackage{array}

\makeatletter

\def\sec{\rm \, sec}

\def\tt{}

\lefthead{A.P. Showman et al., Icarus, in press}
\righthead{A.P. Showman et al., Icarus, in press}
\received{February, 2010}

\sectionnumbers
\printfigures                	

\authoraddr{A.P. Showman, Department of Planetary Sciences, Lunar and
Planetary Laboratory, University of Arizona, Tucson, AZ 85721
(showman@lpl.arizona.edu).  Part of the work was completed while APS
was on sabbatical at Columbia
University, Department of Applied Physics and Applied Mathematics,
New York, NY 10027.}    	

\slugcomment{{\it Icarus}, in press}

\begin{document}
\bibliographystyle{ametsoc}

\title{Scaling laws for convection and jet speeds in the giant planets}

\author{Adam P. Showman}
\affil{University of Arizona, Tucson, AZ and Columbia University, New York, NY}
\author{Yohai Kaspi}
\affil{California Institute of Technology}
\author{Glenn R. Flierl}
\affil{Massachusetts Institute of Technology}

\begin{abstract}
Three dimensional studies of convection in deep spherical shells have
been used to test the hypothesis that the strong jet streams on Jupiter,
Saturn, Uranus, and Neptune result from convection throughout the
molecular envelopes. Due to computational limitations, these simulations
must be performed at parameter settings far from Jovian values and
generally adopt heat fluxes 5--10 orders of magnitude larger than the
planetary values. Several numerical investigations have identified trends for
how the mean jet speed varies with heat flux and viscosity, but no
previous theories have been advanced to explain these trends. Here, we
show using simple arguments that if convective release of potential
energy pumps the jets and viscosity damps them, the mean jet speeds
split into two regimes. When the convection is weakly nonlinear, the
equilibrated jet speeds should scale approximately with $F/\nu$, where 
$F$ is the convective heat flux and $\nu$ is the viscosity. When the
convection is strongly nonlinear, the jet speeds are faster and should
scale approximately as $(F/\nu)^{1/2}$. We demonstrate how this regime shift
can naturally result from a shift in the behavior of the jet-pumping
efficiency with heat flux and viscosity. Moreover, both Boussinesq and
anelastic simulations hint at the existence of a third regime where, at
sufficiently high heat fluxes or sufficiently small viscosities, the jet
speed becomes independent of the viscosity. We show based on
mixing-length estimates that if such a regime exists, mean jet speeds
should scale as heat flux to the 1/4 power. 
Our scalings provide a good match to the mean jet speeds obtained in
previous Boussinesq and anelastic,  three-dimensional simulations
of convection within giant planets over a broad range of parameters.
When extrapolated to the real heat fluxes, these scalings suggest that
the mass-weighted jet speeds in the molecular envelopes of the giant
planets are much weaker---by an order of magnitude or more---than the speeds
measured at cloud level.
\end{abstract}

\begin{article}

\section{Introduction}
\label{introduction}

At the cloud levels near $\sim$$1\,$bar pressure, numerous east-west (zonal) 
jet streams dominate the meteorology of the giant planets Jupiter, Saturn, 
Uranus, and Neptune, but the depth to which these jets extend into the 
interior remains unknown.   Endpoint theoretical scenarios range from 
weather-layer models where the jets are confined to a layer several scale 
heights deep to models where the jets extend throughout the molecular envelope 
($\sim$$10^4\,$km thick) on cylinders parallel to the rotation axis
\citep[for a review see][]{vasavada-showman-2005}.  For Jupiter,
in-situ observations by the Galileo probe at 
$7^{\circ}$N latitude show that the equatorial jet extends to at least 
$\sim$20 bars ($\sim$150$\,$km below the visible clouds) 
\citep{atkinson-etal-1997}, and indirect inferences suggest that the jets 
at other latitudes extend to at least $\sim$5--10 bars pressure
\citep[e.g.,][]{dowling-1995b, morales-juberias-dowling-2005,
legarreta-sanchez-lavega-2008, sanchez-lavega-etal-2008}.  At Neptune, 
gravity data suggest that the fast jets are confined to the outermost 
few percent of the planet's mass \citep{hubbard-etal-1991}.   
Comparable data are currently lacking 
for Jupiter and Saturn but will be obtained by NASA's {\it Juno}
and {\it Cassini} missions, respectively, in coming years.

Three-dimensional (3D) numerical simulations of convection in rotating 
spherical shells have been performed by several groups to investigate 
the possibility that the jets on the giant planets result from 
convection in the interior.   
Both free-slip and no-slip momentum boundary conditions at the inner 
and outer boundaries have been explored;
of these, the free-slip case---which allows the development of 
strong zonal flows---is most relevant to giant planets.  So far,
such studies have neglected the high electrical conductivity and
coupling to magnetic fields expected to occur in the deep 
($\gtrsim 1\,$Mbar) planetary interior.

In this 
line of inquiry, most studies to date make the Boussinesq assumption in which 
the basic-state density, thermal expansivity, and other background 
properties are assumed constant with planetary radius; the convection
is driven by a constant temperature difference imposed between the bottom
hot boundary and top cold boundary \citep{aurnou-olson-2001, christensen-2001, 
christensen-2002, aurnou-heimpel-2004, heimpel-etal-2005,
heimpel-aurnou-2007, aurnou-etal-2008}.   These studies show that 
convection with free-slip boundaries
can drive multiple jets with speeds greatly exceeding the 
convective speeds.  {\tt Studies using thick shells tend to produce
$\sim3$--5 jets \citep{aurnou-olson-2001, christensen-2001, christensen-2002}.
When the shell thickness is only $\sim$10\% of the planetary radius, 
then at least under some parameter combinations, $\sim$15--20 jets can occur, 
similar to the number observed on Jupiter and Saturn \citep{heimpel-etal-2005, 
heimpel-aurnou-2007}.  Nevertheless, many factors in addition to shell thickness
can affect the number of jets.}

In real giant planets, 
the density and thermal expansivity each vary by several orders of magnitude 
from the cloud layer to the deep interior, and a new generation of 
convection models is emerging to account for this strong radial
variation in basic-state properties.  Using the anelastic approximation,
which accounts for this layering, \citet{evonuk-glatzmaier-2006,
evonuk-glatzmaier-2007}, \citet{evonuk-2008} and
\citet{glatzmaier-etal-2009} present idealized two-dimensional (2D) 
simulations in the equatorial plane exploring hypothetical 
basic-state density profiles, with density varying by up to a 
factor of 55 across the convection zone.  In contrast, 
\citet{jones-kuzanyan-2009} present 3D simulations using an idealized
basic-state density structure, with density varying by a factor
of up to 148, while \citet{kaspi-etal-2009, kaspi-etal-2010} 
present 3D simulations
with a realistic Jovian interior structure, with density varying by
nearly a factor of $10^4$ from the deep interior to the 1-bar level.  
These anelastic studies likewise suggest that 
the jets could penetrate deeply through the molecular envelope.

\begin{figure}
\includegraphics[scale=0.45]{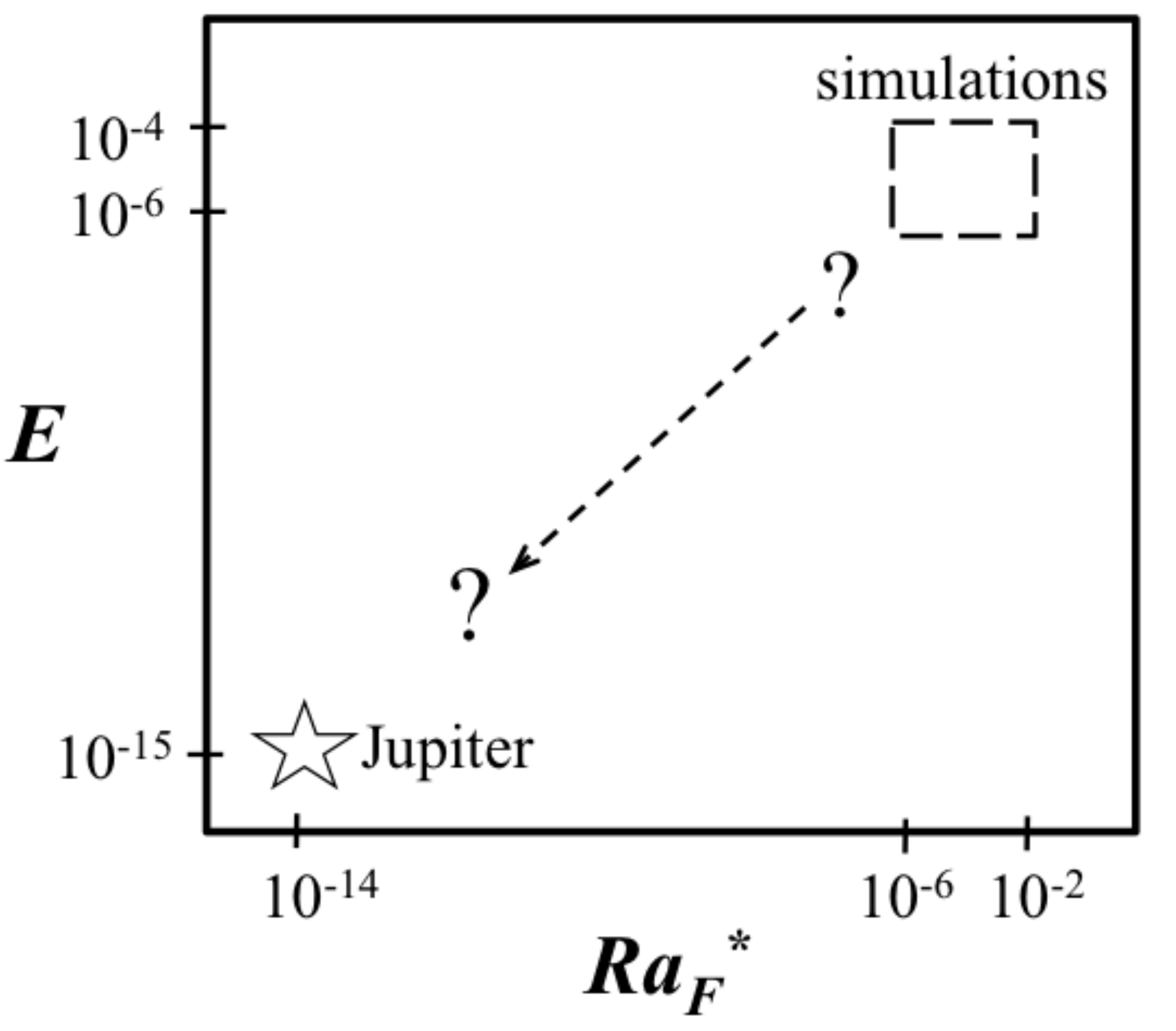}
\caption{Illustration of the parameter space associated with convection
in rotating spherical shells.  The modified-flux Rayleigh number $Ra_F^*$ 
(abscissa) and Ekman number $E$ (ordinate) can be 
viewed as non-dimensionalized 
heat flux and viscosity, respectively.  The Prandtl number, giving 
the ratio of viscosity to thermal diffusivity, constitutes a third
dimension.  Published simulations access a small
region of parameter space with values of $Ra_F^*$ and $E$ typically
$10^{-6}$ or greater, but the actual values for Jupiter are $\sim$$10^{-14}$
or less.  It is unknown whether convection at Jupiter-like values of
$Ra_F^*$ and $E$ would produce Jupiter-like wind speeds.}
\label{schematic}
\end{figure}

A challenge with all of the above-described simulations is that, 
for computational
reasons, they must be performed using heat fluxes and viscosities that 
differ greatly from those on Jupiter, Saturn, Uranus, and Neptune 
{\tt (Fig.~\ref{schematic}).}
Thus, while simulations can produce jets with
speeds similar to the observed values of $\sim$100--$200\rm\,m\,sec^{-1}$
for some combinations of parameters
\citep[e.g.,][]{christensen-2001, christensen-2002, aurnou-olson-2001,
heimpel-etal-2005, heimpel-aurnou-2007, aurnou-etal-2007, aurnou-etal-2008,
kaspi-etal-2009, jones-kuzanyan-2009},
this does not imply that convection in the interior of real giant 
planets would necessarily produce jets with such speeds.  In fact, depending on
the parameter combinations, simulations with free-slip
boundary conditions can produce
jets that equilibrate to mean speeds\footnote{Calculated in a 
suitably defined way, such as $\sqrt{2K/M}$, 
where $K$ and $M$ are the total kinetic energy and mass; 
this is essentially the mass-weighted characteristic wind speed of 
the fluid.} ranging over {\tt many orders of magnitude,
from arbitrarily small} (less than $1\rm\,m\,sec^{-1}$) 
to $1000\rm\,m\,sec^{-1}$ or more.
Assessing the likely wind speeds in the molecular envelopes of the
real giant planets---and determining whether their observed
jets can be pumped by convection in their interiors---requires 
the development of a theory that can be extrapolated 
from the simulation regime to the planetary regime. 

Currently, however, there is no published theory that can explain
the jet speeds obtained in simulations with free-slip boundaries
nor their dependence on heat flux, viscosity, and other parameters.
Several investigations have presented scaling laws describing 
how the mean zonal-wind speeds vary with control parameters when 
{\it no-slip} boundary conditions are used, as potentially relevant 
to Earth's outer core \citep[e.g.,][]{aurnou-etal-2003, aubert-2005}.
These are essentially theories for the magnitude of wind shear
in the fluid interior for cases where the zonal velocity is 
pinned to zero at the boundaries.  However, these scaling laws are 
not applicable to the giant planets, 
where the fluid at the outer boundary can move freely,
lacks a frictional Ekman layer, and exhibits strong jets. 
Several attempts have also been made to
quantify how the {\it convective} velocities scale with parameters,
but---regardless of the boundary conditions---these relationships 
cannot be applied to the {\it jet} speeds because
the convective and jet speeds can differ greatly
and their ratio may depend on the heat
flux and other parameters.   

The goal of this paper, therefore, is to develop scaling 
laws for how the jet speeds depend on heat flux and viscosity 
that explain the simulated behavior within 
the simulated regime and, ideally, allow an extrapolation to real planets.
The simulations themselves make a number of simplifications
(e.g., ignoring magnetohydrodynamics in the deep interior at 
pressures exceeding $\sim$$1\,$Mbar), but in our view building
a theory of this idealized case is a prerequisite for understanding
more realistic systems.

We first quantify the degree of overforcing in current studies, since
this issue has received little attention in the literature 
(Section~\ref{overforcing}).  Next, we quantify the dependence
of the convective speeds on planetary parameters and
compare them to results from an anelastic general circulation model
from \citet{kaspi-etal-2009} (Section \ref{convective}).  
Armed with this information, 
we construct simple scalings for the characteristic jet speeds in three
regimes.   In the first two regimes (Section 4), the
viscosity is large enough so that viscous damping of the jets
provides the dominant kinetic-energy loss mechanism.  
\citet{christensen-2002} suggested the existence of a third regime
where the jet speeds become independent of the viscosity; we 
construct possible scalings for this regime in Section \ref{asymptotic}.
In Section~\ref{combine}, we combine
the three regimes and discuss extrapolations to Jupiter.   
Section~\ref{conclusions} concludes.

\section{Degree of overforcing}
\label{overforcing}

For numerical reasons, current 3D simulations of convection 
in the giant planets must use viscosities many orders of magnitude 
larger than the molecular viscosities.  This results from the
coarse grid resolution in the models: to be numerically converged, 
such a model must have convective boundary layer and convective 
plume thicknesses of at least a gridpoint, and this requires
very large viscosities.  Given the enhanced damping of the 
kinetic energy implied by this constraint, such simulations can achieve
wind speeds similar to those on the giant planets only by
adopting heat fluxes orders of magnitude too large.  

Moreover, even in the absence of such a viscous damping, 
enhanced heat fluxes are computationally necessary simply
to achieve spin up within available integration times.
To illustrate, imagine
a giant planet that initially has no winds, and ask how long
it would take to spin the zonal jets up to full strength if the
jets penetrate through the molecular region (down to $\sim$$1\,$Mbar
pressure on Jupiter and Saturn).  This corresponds to a kinetic
energy per area of $\sim u^2\Delta p/g$, where $u$ is the characteristic
jet speed ($\sim30\rm \,m\,sec^{-1}$ on Jupiter), $g$ is gravity, and
$\Delta p\sim 1\,$Mbar
is the pressure thickness across which the jets are assumed to extend.
On giant planets, the work to spin up the winds comes from the convection,
driven by {\tt internal heat fluxes ranging from $0.3\rm\,W\,m^{-2}$ or less
for Uranus and Neptune 
to $5\rm\,W\,m^{-2}$ for Jupiter}.  Given these small available fluxes,
the characteristic pumping time of the jets is $\sim$$10^5\,$years.  
This is a lower limit, because it assumes that the efficiency of
converting the heat flux into kinetic energy is close to 100\%;
in reality, the efficiency will be less, and much of the power
produced will be used to resist frictional losses in the simulations.
In contrast, published state-of-the-art 
high-resolution numerical simulations, if 
expressed in dimensional units, typically integrate months to years
\citep[e.g.,][]{christensen-2001, christensen-2002, aurnou-olson-2001, 
aurnou-etal-2008, heimpel-etal-2005, heimpel-aurnou-2007, jones-kuzanyan-2009,
kaspi-etal-2009}.  Thus, the heat fluxes must be 
enhanced by at least 3--5 orders of magnitude simply to allow 
the simulations to spin up within achievable integration times.


Most convection papers cast the equations in nondimensional
form, rendering unclear the actual degree to which these simulations
are overforced.  One can quantify the degree of overforcing
as follows.  In these simulations, the specified control
parameters are the Rayleigh number $Ra$, the Ekman number $E$,
and the Prandtl number $Pr$, defined as follows:

\begin{equation}
Ra={\alpha g \Delta T D^3\over \kappa \nu}, \qquad E={\nu\over \Omega D^2},
\qquad Pr={\nu\over\kappa}
\label{Ra}
\end{equation}
where $\alpha$, $g$, $\kappa$, and $\nu$ are the
thermal expansivity, gravity, thermal diffusivity, and kinematic
viscosity. $D$ is the thickness of the convecting layer, $\Delta T$
is the imposed temperature drop across this layer, and $\Omega$
is the planetary rotation rate.  

The heat flux, which is an output, is typically cast
as a Nusselt number $Nu$, which gives the ratio of the total (convective
plus conductive) heat flux to a reference conductive flux that
would be conducted across a system with the same temperature drop and
thickness:
\begin{equation}
Nu={F_{\rm tot}\over F_{\rm cond}} = {D F_{\rm tot}\over \kappa \rho c_p \Delta T},
\label{Nu}
\end{equation}
where $F_{\rm tot}$ is the total flux, the thermal conductivity is 
$\kappa\rho c_p$, and  $F_{\rm cond}=\kappa\rho c_p \Delta T/D$.
At Rayleigh numbers far exceeding the critical value, the total heat flux approximately 
equals the convective heat flux.

So how may we determine the {\it dimensional} heat flux 
for a nondimensional simulation with specified $Ra$, $E$, $Pr$, 
and $Nu$?  Using (1) and (2) shows that
\begin{equation}
F_{\rm tot}={\Omega^3 D^2 \rho c_p\over \alpha g} {E^3 Ra Nu\over Pr^2}.
\label{overforcing1}
\end{equation}

Because convection is driven by buoyancy associated with a heat
flux, an alternate Rayleigh number, defined in terms of the heat flux
rather than a temperature contrast,
is useful:
\begin{equation}
Ra_F = {\alpha g F_{\rm tot} D^4\over \rho c_p \nu \kappa^2}.
\label{Ra_F}
\end{equation}

\citet{christensen-2002} argued that, in the limit of small diffusivities,
the fluid behavior should become independent of the diffusivities,
and he defined a modified heat-flux Rayleigh number that is
independent of these diffusivities.  This can be constructed
by multiplying Eq.~(\ref{Ra_F}) by $E^3 Pr^{-2}$, 
giving\footnote{\citet{christensen-2002} made alternate definitions 
of $Ra^*_F$ and $Nu$ that included a non-dimensional factor of $\eta$ 
(the ratio of the inner-to-outer radius adopted in the simulations) 
in the numerator of Eqs.~\ref{Nu} and \ref{Ra_F}. We forgo this
factor here.}

\begin{equation}
Ra_F^* = {\alpha g F_{\rm tot}\over \rho c_p \Omega^3 D^2}.
\label{Ra_Fstar}
\end{equation}
In the limit of small diffusivities,
one thus expects that the behavior should scale with $Ra_F^*$
rather than $Ra$.   Casting Eq.~(\ref{Ra_Fstar}) as an expression for 
heat flux, we have
\begin{equation}
F_{\rm tot}={\Omega^3 D^2 \rho c_p\over \alpha g} Ra_F^*.
\label{overforcing2}
\end{equation}

For a giant planet, all the dimensional
parameters in Eq.~(\ref{overforcing2}) are known.  Consider Jovian values of 
$\Omega=1.74\times10^{-4}\sec^{-1}$, $D\approx 2\times 10^4\rm\,km$, 
$c_p=1.3\times10^4\rm\,J\,kg^{-1}\,K^{-1}$, and $g=26\rm\,m\,sec^{-2}$.
For interpreting Boussinesq 
simulations in the context of giant planets, an ambiguity is that
the thermal expansivity and density vary greatly from atmospheric values 
at the top to liquid values in the interior.  Using atmospheric values
corresponding to a pressure of $\sim$$100\,$bars, where the temperature is
$\sim$$600\rm\,K$, we have $\rho\approx 4\rm\,kg\,m^{-3}$ 
and $\alpha\approx 1.6\times10^{-3}\rm\,K^{-1}$.  Considering the 
simulation in \citet{christensen-2001}, where $Pr=1$, $E=3\times10^{-5}$,
$Ra=10^8$, and $Nu\approx10$,
we then have heat fluxes of $\sim$$10^5\rm\,W\,m^{-2}$.  However, given
that the vast bulk of the interior behaves as a liquid, much more
appropriate values are $\rho\sim10^3\rm\,kg\,m^{-3}$ and 
$\alpha\sim10^{-5}\rm\,K^{-1}$, relevant to the bulk of Jupiter's interior.  
These values imply fluxes of $\sim$$10^9$--$10^{10}\rm\,W\,m^{-2}$.
Similar estimates imply that the simulations of \citet{heimpel-aurnou-2007} and 
\citet{aurnou-etal-2008} adopt a heat flux of $\sim$$10^8\rm\,W\,m^{-2}$.
\citet{aurnou-etal-2007}'s simulations of Uranus and Neptune adopt 
a heat flux of $\sim$$10^7\rm\,W\,m^{-2}$.

Given the great variation of density and thermal expansivity with
radius on a giant planet, future studies will increasingly adopt
models that incorporate the full radial variation of these parameters.
This will require extending the definition of Rayleigh number. 
We adopt the following definition, in which a mass-weighted average
is performed over the radially varying quantities \citep{kaspi-etal-2009}:
\begin{equation}
Ra_F^* = {1\over \Omega^3 D^2}\langle {\alpha g F_{\rm tot}\over \rho c_p}\rangle,
\label{Ra_Fstar-integrated}
\end{equation}
where the mass-weighted average is defined as $\langle ...\rangle=
 \int (...) \rho r^2\,dr/\int \rho r^2\,dr$.

Anelastic simulations that adopt realistic radial profiles for
the density and thermal expansivity lack the ambiguity described
above.  \citet{kaspi-etal-2009} force the flow not through a 
constant-temperature boundary condition but by introducing 
internal heating and cooling.
Integrating their heating profile over the depth of the planet gives an
effective heat flux with a maximum of $\sim$$10^9\rm\,W\,m^{-2}$
in the deep interior with values decaying to zero in the atmosphere. 
Similarly, \citet{jones-kuzanyan-2009} estimate a dimensional heat
flux of $\sim$$10^9\rm\,W\,m^{-2}$ for their simulations, 
broadly consistent with the estimates made here.

The implication is that the simulations in \citet{christensen-2001, 
christensen-2002}, \citet{aurnou-olson-2001}, \citet{heimpel-aurnou-2007},
\citet{aurnou-etal-2008}, \citet{kaspi-etal-2009} and related studies
are overforced by $\sim$6--10 orders of magnitude. 
The overforcing compensates for the numerical need to use high diffusivities 
and achieve a steady state over reasonable time scales.  In non-dimensional
terms, the numerical models adopt $Ra_F^*$ of $10^{-8}$ or greater
while Jupiter's $Ra_F^*$ value is $\sim$$10^{-14}$.

\section{Convective velocities and buoyancies}
\label{convective}
In a giant planet, convection in the interior transports the interior 
heat flux.  To order-of-magnitude, one thus expects that
\begin{equation}
F\sim \rho c_p w \delta T,
\label{heat-transport}
\end{equation}
where $F$ is the convected heat flux, $w$ is the characteristic
magnitude of the vertical convective velocity, and $\delta T$ is
the characteristic magnitude of the temperature difference between
a convective plume and the surrounding fluid.

In a rapidly rotating, low-viscosity fluid, the scaling for  
convective velocities is typically written \citep[e.g.,][]{stevenson-1979,
golitsyn-1980, golitsyn-1981, boubnov-golitsyn-1990, 
fernando-etal-1991}\footnote{Equation~(\ref{vert-velocity})
can be heuristically derived by assuming that the convective buoyancy 
force per mass, $g\alpha\,\delta T$, is balanced by the vertical Coriolis
force due to the turbulent eddy motions, $\Omega u'$, where $u'$
is the horizontal eddy velocity.  Assuming that the turbulent eddy
motions are approximately isotropic ($u'\sim w$) and invoking
Eq.~(\ref{heat-transport}) then leads to Eq.~(\ref{vert-velocity}).}
\begin{equation} 
w \sim \left({\alpha g F\over \rho c_p \Omega}\right)^{1/2}.
\label{vert-velocity}
\end{equation}
It is known that Eq.~(\ref{vert-velocity}) provides
a reasonable representation of convective velocities for 
laboratory experiments in rotating tanks 
\citep[e.g.,][]{golitsyn-1981, boubnov-golitsyn-1990, fernando-etal-1991}.
Several authors have also suggested that Eq.~(\ref{vert-velocity})
is relevant for the dynamo-generating regions of planetary interiors,
where a three-way balance between buoyancy, Coriolis, and
Lorentz forces is often assumed \citep[e.g.,][]{starchenko-jones-2002,
stevenson-2003}.  To our knowledge, however, no broad assessment
of its accuracy has been made for convection in the molecular 
interior of a giant planet.  To do so, we performed three-dimensional
numerical simulations of convection in Jupiter's interior using
an anelastic general circulation model based on the MITgcm
\citep{kaspi-etal-2009}.  The radial dependence of gravity, basic-state density,
compressibility, and thermal expansivity correspond to a realistic Jupiter
interior structure calculated with the SCVH equation 
of state \citep{saumon-etal-1995}.  
Simulations with a broad range of parameters were explored 
(Table~\ref{symbol-key}).  {\tt All of our simulations are in a rapidly rotating,
low-viscosity regime, with geostrophic balance holding on large scales 
\citep[see, e.g.,][Fig.~6]{kaspi-etal-2009}.}

\begin{figure}
\includegraphics[scale=0.45]{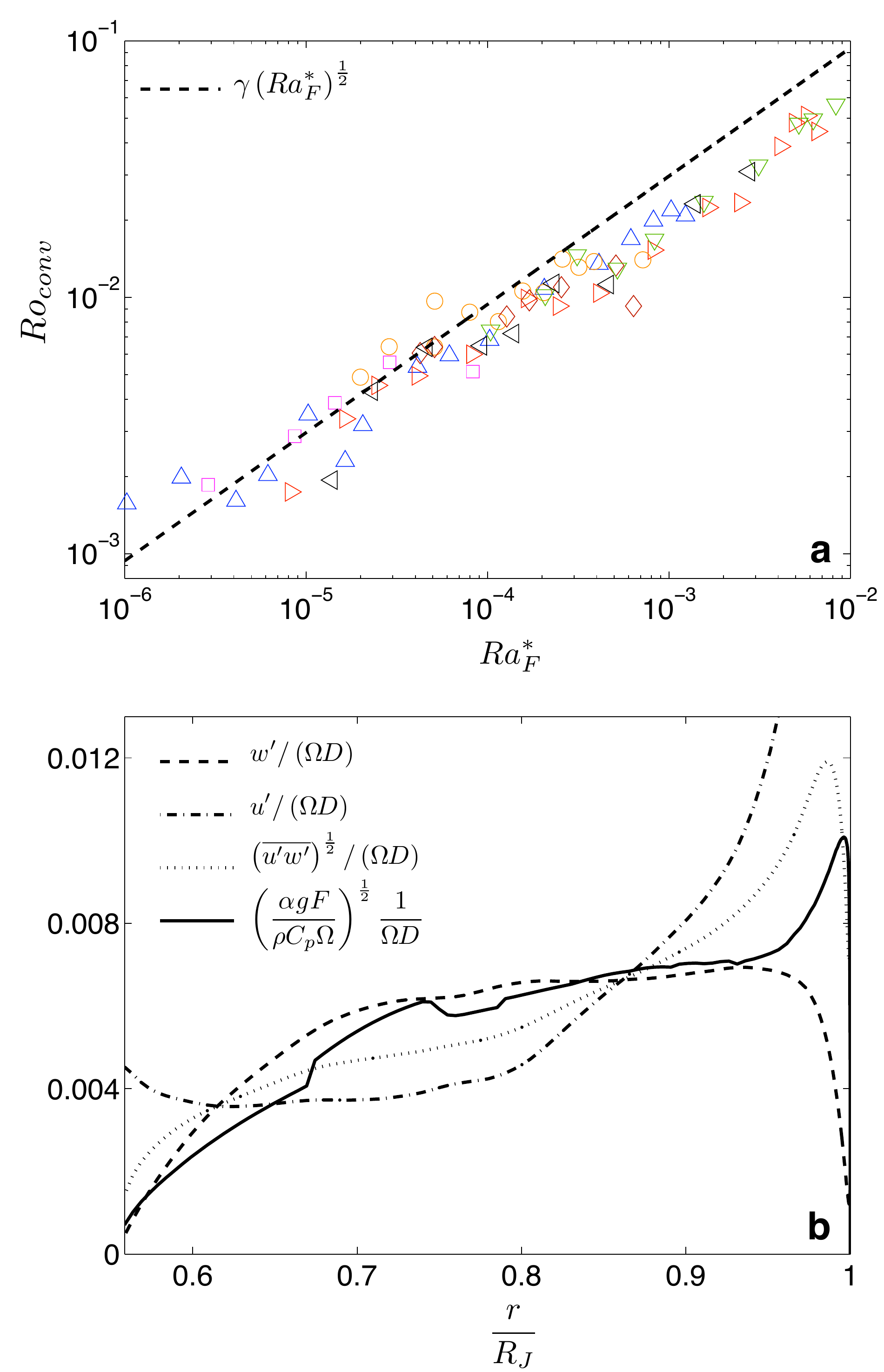}
\caption{{\it Top:} Mass-weighted convective vertical wind speed (calculated
as the deviation of vertical wind speed from its zonal mean)
versus modified heat-flux Rayleigh number for a range of anelastic simulations.
The velocities are expressed as a convective Rossby number, defined as
velocity divided by $\Omega D$.  Dashed line is the scaling of Eq.~(\ref{vert-velocity-nondim}) with $\gamma=1$.  
See Table~\ref{symbol-key} for definitions of the symbols.
{\it Bottom:} Radial variation of the horizontally averaged velocity
components for {\tt an anelastic simulation with parameters $Ra_F^*=2.89\times10^{-5}$,
$E=1.5\times10^{-4}$, and $Pr=10$}.  Depicts deviation of
vertical velocity from its zonal mean
(dashed), deviation of the zonal velocity from its zonal mean (dashed-dotted),
the square root of their zonally averaged correlation 
$(\overline{u'w'})^{1/2}$ (dotted), and the scaling from 
Eq.~(\ref{vert-velocity}) evaluated using the radially varying
thermal expansivity $\alpha$, basic-state density $\tilde{\rho}$,
heat flux $F$, and gravity.  {\tt Note that all the anelastic simulations in
this paper use the full Jovian interior structure with a factor of $\sim$$10^4$
variation in density from top to bottom.}
For more details on the experiments see
\citet{kaspi-2008} and  \citet{kaspi-etal-2009}.}
\label{w-anelastic}
\end{figure}

\begin{table*}
\includegraphics[scale=0.9]{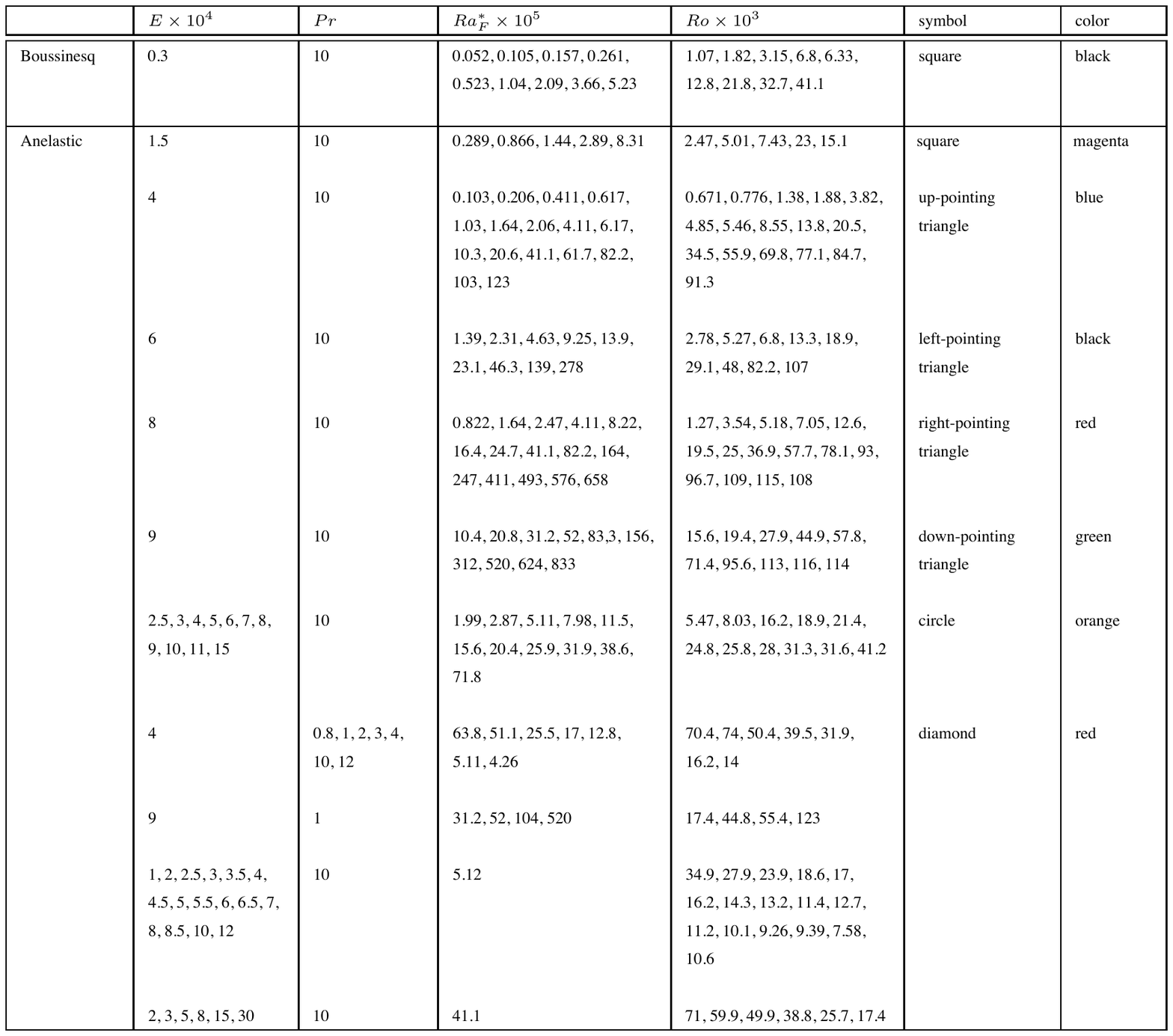}
\vskip -290pt 
\caption{List of numerical integrations, using the model of 
\citet{kaspi-etal-2009}, that are presented in this paper.  Each
row lists a sequence of simulations.  Columns 2, 3, and 4 give
the control parameters, and column 5 gives the global-mean,
mass-weighted Rossby number.  The last two columns give the
symbol type and color used to present each sequence of simulations
in Figs.~\ref{w-anelastic}, \ref{weakly-strongly}, and \ref{rossby-rayleigh}. 
All the anelastic model integrations in this paper adopt the full Jovian radial interior structure
of gravity, density, thermal expansivity, compressibility, and other thermodynamic 
properties from \citet{kaspi-etal-2009}, corresponding to a factor of $\sim$$10^4$ variation of 
density from top to bottom.}
\label{symbol-key}
\end{table*}

Our simulations show that Eq.~(\ref{vert-velocity}) provides a
good approximation over a wide range of heat fluxes and rotation rates.  
Figure~\ref{w-anelastic} shows the mass-weighted mean convective 
velocities from the \citet{kaspi-etal-2009} anelastic simulations 
(presented as a convective Rossby number, namely convective vertical velocity
divided by $\Omega D$) versus modified-flux Rayleigh number and compares it to 
Eq.~(\ref{vert-velocity}). 
For the vast majority of simulations, the simulations differ
less than a factor of $\sim$3 from the scaling, indicating that 
Eq.~(\ref{vert-velocity}) is quite accurate.  This analysis
complements \citet{kaspi-etal-2009}'s finding that, within a 
specific simulation, Eq.~(\ref{vert-velocity}) captures the 
approximate {\it radial dependence} of the convective velocity 
across the interior, over which $\alpha/\rho$ varies by a factor of 
$\sim10^5$ (Fig.~\ref{w-anelastic}, {\it bottom}).  
Together, these results suggest that Eq.~(\ref{vert-velocity})
is valid not only for the mass-weighted mean velocities but {\it locally}
within the fluid.

Laboratory experiments suggest that a prefactor should exist on the
right side of Eq.~(\ref{vert-velocity}) with a value close to 2 
\citep{golitsyn-1981, boubnov-golitsyn-1990, fernando-etal-1991}; 
however, our simulations are better matched when no prefactor is used.

We can nondimensionalize the convective velocities using a 
convective Rossby number, $Ro_{\rm conv} \equiv w/(\Omega D)$.
The convective flux, $F$, equals $F_{\rm tot}$ minus the conductive
flux.  The conductive flux is only important near the critical
Rayleigh number, and in this limit it becomes $F_{\rm cond}$
from Section~\ref{overforcing}.  Making the approximation that
the conductive flux equals $F_{\rm cond}$, Eq.~(\ref{vert-velocity})
nondimensionalizes to yield
\begin{equation}
Ro_{\rm conv} \sim \gamma (Ra_F^* - Ra_F^{*\rm crit})^{1/2},
\label{vert-velocity-nondim}
\end{equation}
where $Ra_F^{*\rm crit}$ is the critical modified flux Rayleigh
number for the onset of convection.  Thus, at the critical Rayleigh
number, the expression properly predicts zero convective velocities,
and at greatly supercritical Rayleigh numbers it predicts $Ro_{\rm conv}
\sim\gamma (Ra_F^*)^{1/2}$.  
We have introduced a dimensionless prefactor $\gamma$ 
for generality, but consistent with Fig.~\ref{w-anelastic}, 
we will assume $\gamma=1$ when performing numerical estimates.

Like Eq.~(\ref{vert-velocity-nondim}),
many of our subsequent scalings will depend on the difference between
the actual and critical Rayleigh numbers.  For notational brevity,
we therefore define this difference as
\begin{equation}
Ra_F^{\Delta} \equiv Ra_F^*-Ra_F^{*\rm crit}.
\label{RaFdelta}
\end{equation}

Equations~(\ref{heat-transport}--\ref{vert-velocity}) imply that
the fractional density associated with convective plumes should
scale as
\begin{equation}
\alpha \delta T\sim \left({F \Omega\alpha \over \rho c_p  g}\right)^{1/2}.
\label{plume-buoyancy}
\end{equation}

\citet{aubert-etal-2001} proposed an alternate scaling for the convective
velocities in the limit of negligible viscosities, 
$Ro_{\rm conv}\sim (Ra_F^*)^{2/5}$.  They performed rotating
laboratory experiments in liquid gallium and water, and showed that this
expression provides a reasonable fit to the convective velocities inferred
for their experiments.  In the numerical simulations of 
\citet{christensen-2002} and \citet{christensen-aubert-2006}, the 
poloidal component of the velocity field (which includes the convective 
velocities) depends on the Ekman number, but at small Ekman number
seems to be converging toward an asymptotic dependence that is
reasonably well represented by this 2/5 scaling (with a prefactor
of 0.5).  In our case, Eqs.~(\ref{vert-velocity})--(\ref{vert-velocity-nondim})
provides a slightly better fit to the convective velocities, 
although the 2/5 scaling (with a prefactor) is also
adequate.  Conversely, overplotting 
Eqs.~(\ref{vert-velocity})--(\ref{vert-velocity-nondim}) against
\citet{christensen-2002}'s simulation data shows that 
they match essentially as well as the 2/5 scaling.

\section{Scaling for the jet speeds}
\label{viscous}

\subsection{Experimental data}
\label{sim-data}

{\tt Our goal is to understand the physical processes governing the 
global-mean jet speeds as a function of heat flux and viscosity
(i.e., Rossby number as a function of Ekman and Rayleigh numbers)
for low-viscosity, rapidly rotating convection in spherical
shells.  To characterize this functional dependence requires
numerous ($\sim$100 or more) numerical integrations so
that the available parameter space is adequately sampled.  While
it is numerically possible in three-dimensional simulations to reach Ekman numbers as low
as 3--$4\times10^{-6}$ \citep[e.g.,][]{heimpel-etal-2005, heimpel-aurnou-2007,
aurnou-etal-2008, jones-kuzanyan-2009}, this requires intensive computation and 
precludes the exhaustive survey of parameter space required to develop
scaling laws. Instead, our strategy is to adopt more modest Ekman
numbers (extending to $10^{-4}$ to $10^{-5}$), which ensure that the dynamics
are still in the low-viscosity, rapidly rotating regime (with geostrophic
balance holding on large scales) yet allow numerous simulations to be 
performed.  Here we present over 100 simulations.  In addition, we extensively
use the dataset of \citet{christensen-2002}, who systematically characterized
the global-mean Rossby numbers as a function of Ekman and Rayleigh numbers
in Boussinesq simulations with Ekman numbers as low as $10^{-5}$.  Despite being
far from Jovian parameter values, all the simulations we present are in the appropriate
rapidly rotating, low-viscosity regime, with geostrophic balance holding at large
scales.}

\begin{figure}
\includegraphics[scale=0.4]{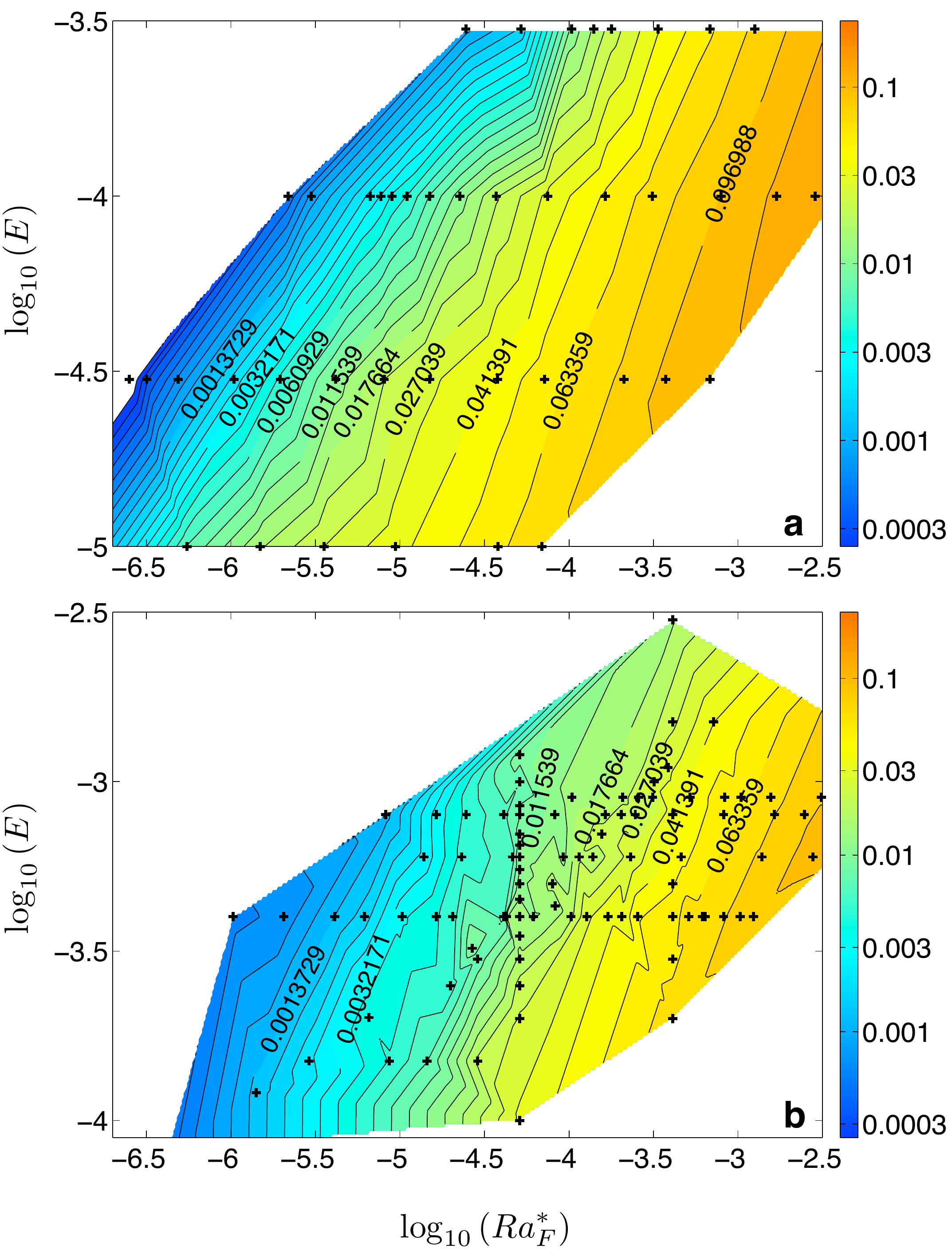}
\caption{Dependence of mass-weighted, domain-averaged jet speed on parameters
from published simulations.  Depicts the Rossby number (color and contours)
as a function of modified flux Rayleigh number (abscissa) and Ekman number 
(ordinate) for Boussinesq simulations from \citet{christensen-2002} (top) 
and anelastic simulations as described in Table~\ref{symbol-key} using
the model of \citet{kaspi-etal-2009} (bottom).  Note that contours are evenly
spaced in $\log(Ro)$ and that all figures in this paper use identical
contour values and colorbars, facilitating intercomparison.  
In both panels, the plusses
denote the locations of individual simulations.  The range
of Ekman numbers plotted is different in the two cases; anelastic cases
are computationally more demanding and hence were run at larger
Ekman numbers than the Boussinesq cases.  {\tt The small-scale structure
results from interpolation onto a fine grid of the coarsely spaced 
experimental points and may not be robust.}  The robust feature
is the overall trend of positive slopes of constant-$Ro$ contours, with 
mean slopes close to one, and $Ro$ values ranging from $\sim$0.1 on the 
right to $\sim$0.0001 on the left.  Note that, in both sets of
simulations,  the contours are widely spaced at Rossby numbers exceeding 
$\sim$0.02 (right side of plot) but become tightly spaced at Rossby numbers
less than $\sim$0.02 (left side of plot), suggesting a regime transition.}
\label{rossby-2D-sims}
\end{figure}

These numerical experiments demonstrate that the mean jet speeds depend significantly
on the control parameters $Ra_F^*$ and $E$.  Figure~\ref{rossby-2D-sims}
depicts the global-mean, mass-weighted Rossby number$ Ro\equiv U/(\Omega D)$, where
$D$ is the thickness of the model domain and $U$ is the domain-mean, mass-weighted
wind speed, as a function of the control parameters in the Boussinesq simulations from
\citet{christensen-2002} (top panel) and the anelastic simulations described in
Table~\ref{symbol-key} using the model of \citet{kaspi-etal-2009}
(bottom panel).  The domain-averaged jet speeds range across three orders of magnitude
even within the limited parameter space explored in these models,
with some parameter combinations producing Jupiter-like speeds and others not. 
Rossby numbers range from $\sim$0.0001 to 0.1,
corresponding to domain-mean wind speeds of $\sim$1 to $1000\rm\,m\,sec^{-1}$.
In both cases, larger $Ra_F^*$ and smaller $E$
promote faster wind speeds.  This makes sense qualitatively because,
for a given rotation rate and planetary size, larger $Ra_F^*$
implies larger heat flux  (i.e., stronger forcing of the flow), while
smaller $E$ implies smaller viscosity (i.e., weaker damping of the flow).
\citet{kaspi-etal-2009} found that constant-wind-speed contours exhibit
a slope (in the $\log(Ra_F^*)$--$\log(E)$ diagram) of approximately 5/4.  
The data from \citet{christensen-2002} plotted in Fig.~\ref{rossby-2D-sims} 
likewise indicate that the wind-speed contours in his case 
exhibit a mean slope of approximately 1.  
For both sets of simulations, Fig.~\ref{rossby-2D-sims} also shows 
that the contours tend to be widely spaced toward the right and tightly spaced
toward the left, with an approximate transition at $Ro\approx0.02$.
This suggests a transition between two regimes.

{\tt The processes that determine how mean jet speeds
depend on $Ra_F^*$ and $E$ are not understood, however, and so the trends
in Fig.~\ref{rossby-2D-sims} remain unexplained.    Developing such an 
understanding is important because rotating spherical-shell convection is
an inherently interesting physics problem, and also because
extrapolation into the jovian parameter regime
can only be performed once a theory for the $Ra_F^*$ and $E$-dependence
of the mean jet speeds has been developed.
In the absence of such a theory,  it is unknown whether convection at 
Jupiter-like values of $Ra_F^*$ and $E$ would generate Jupiter-like 
jet speeds.   Explaining Fig.~\ref{rossby-2D-sims}
is the core goal of this paper.}


\subsection{Regime I: Strongly nonlinear regime}
\label{energetics-approach}

 Here, we construct a simple scaling theory based on energetic arguments 
to attempt an explanation for the jet speeds in the regime of 
fast jets ($Ro\gtrsim 0.02$ in Fig.~\ref{rossby-2D-sims}, corresponding
to jet speeds exceeding $\sim100\rm\,m\,sec^{-1}$), which we call 
Regime I.

First consider the forcing.  A convecting fluid parcel traveling at
a vertical speed $w$ releases potential energy per mass at a rate
$\dot P\sim g w \delta \rho/\rho$, where $\rho$ is density and 
$\delta\rho$ is the magnitude of the density contrast between 
plumes and the background environment.  Using the equation of
state $\delta\rho = \alpha\rho \delta T$, where $\rho$ represents
a background density and $\delta T$ is
the magnitude of the temperature contrast between plumes and the
background environment, we can write $\dot P \sim g w \alpha \,\delta T$.
This potential energy is primarily converted into convective kinetic energy.
Some fraction $\epsilon$ of this energy is used to pump the zonal jets;
we provisionally assume $\epsilon$ is constant but return to this assumption
in Section 4{\it c}.
Integrating over the planetary mass, the total power (in W) pumped 
into the jets is then approximately
\begin{equation}
\dot P_{\rm tot} \sim 4 \pi \epsilon \int \alpha \delta T g w \rho r^2 \,dr,
\label{integrated-forcing}
\end{equation}
where the mass of an infinitesimal spherical shell of radial thickness $dr$ is 
$4\pi \rho r^2 \,dr$.  Both 
$\delta T$ and $w$ are {\it a priori} unknown, but they are related
by the fact that convection transports the planet's interior heat flux
(Eq.~\ref{heat-transport}), which implies that $w\, \delta T\sim F/(\rho c_p)$.
Thus, the power available to pump the jets can be written
\begin{equation}
\dot P_{\rm tot} \sim 4\pi \epsilon \int {\alpha g F\over c_p}r^2 \,dr.
\label{integrated-forcing2}
\end{equation}

Energy loss occurs through friction, which we assume acts as a diffusive
damping of the winds with a viscosity $\nu$ (for the simulations,
this would be the model viscosity that enters the definition
of the Rayleigh and Ekman number; see Eq.~\ref{Ra}).
The power per mass dissipated by viscosity is approximately 
$\nu \nabla^2 u^2\sim \nu k^2 u^2$, where $k$ is the wavenumber
of the structures dominating the dissipation,
that is, the dominant wavenumber of $\nabla^2 u^2$.  
This implies a total rate of kinetic energy loss given by
\begin{equation}
\dot{K}_{loss} \sim 4\pi \int \nu k^2 u^2 \rho r^2\,dr.
\label{integrated-loss}
\end{equation}

Now we equate energy gain (Eq.~\ref{integrated-forcing2}) and 
loss (Eq.~\ref{integrated-loss}) to obtain an expression for
mean jet speed.  In the case of a Boussinesq fluid, we note that $\alpha$ 
and $\rho$ are constants and that $F$, $g$, $c_p$, and $u$ are approximately
constant (within a factor of $\sim$3) and can be pulled out of the integral,
yielding\footnote{\tt A similar equation was derived, albeit with different
arguments, by \citet[][Eq.~20]{ingersoll-pollard-1982}.}
\begin{equation}
u\sim k^{-1}
\left({\alpha g \epsilon F\over \rho c_p \nu}\right)^{1/2}
\label{u-viscous-boussinesq}
\end{equation}
The case of an anelastic fluid is complicated by the large
variation of $\alpha$ and $\rho$ with radius, and even $u$
may vary significantly from the top to the bottom of the domain
\citep{kaspi-etal-2009}.  However, in the \citet{kaspi-etal-2009}
simulations, the mean jet speeds are relatively constant (within
a factor of $\sim3$) throughout most of the domain; only the top
few \% of the fluid mass experiences significant wind shear along
the Taylor columns.  To obtain a crude expression for the mass-weighted
mean jet speed in the molecular region, we therefore assume that $u$ is
constant, allowing us to write:
\begin{equation}
u\sim k^{-1} \left({\epsilon\over\nu}\right)^{1/2}\left[
{\int {\alpha g F\over \rho c_p}\rho r^2\,dr \over \int \rho r^2\,dr}
\right]^{1/2}.
\label{u-viscous-anelastic}
\end{equation}

Let us nondimensionalize these expressions so that they may
be compared with the simulation results shown in Fig.~\ref{rossby-2D-sims}.
For both cases, we use $Ro\equiv u/(\Omega D)$ and the definition of
$E$ given in Eq.~(\ref{Ra}).  Using the appropriate definition
of the modified flux Rayleigh number yields the same nondimensional
expression for both the Boussinesq and anelastic cases. Again approximating 
the convective flux $F$ as the total flux minus $F_{\rm cond}$, we
obtain
\begin{equation}
Ro\approx{\epsilon^{1/2} \over k D} \left({Ra_F^{\Delta}\over E}\right)^{1/2},
\label{rossby-viscous}
\end{equation}
where $Ra_F^{\Delta}\equiv Ra_F^* - Ra_F^{*\rm crit}$ (see Eq.~\ref{RaFdelta}).
In Eq.~(\ref{rossby-viscous}), $Ra_F^*$ is given by Eq.~(\ref{Ra_Fstar}) 
for the Boussinesq case and by Eq.~(\ref{Ra_Fstar-integrated}) for 
the anelastic case.  Note that, in the highly supercritical regime
in which we expect Eq.~(\ref{rossby-viscous}) to be valid, the critical
Rayleigh number is generally negligible, so
that $Ra_F^{\Delta}$ approximately equals $Ra_F^*$.

\begin{figure}
\includegraphics[scale=0.4]{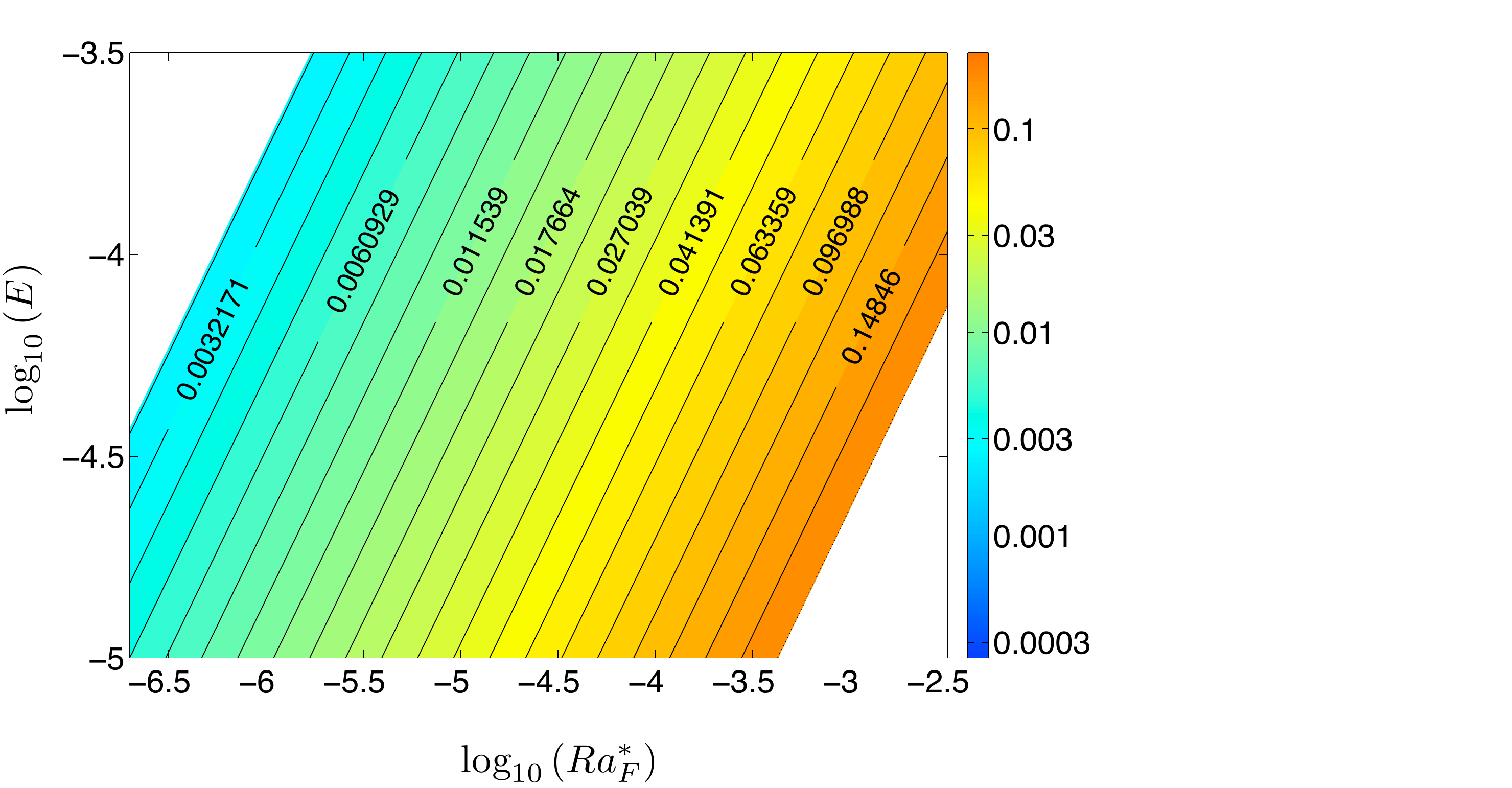}
\caption{Prediction of the scaling law for Regime I given in 
Eq.~(\ref{rossby-viscous}), in which the jet-pumping efficiency
$\epsilon$ is assumed constant.  Depicts contours of Rossby number 
versus modified flux Rayleigh number (abscissa) and Ekman number (ordinate).
Adopts $kD=5\pi$ and $\epsilon=0.3$ (see text).
Note the similarity of the slopes and values of the 
contours with the simulation results plotted in Fig.~\ref{rossby-2D-sims},
particularly for Rossby numbers greater than $\sim$0.02.}
\label{rossby-2D-viscous}
\end{figure}

Thus, this simple theory predicts that the characteristic jet
speed is proportional to $(F/\nu)^{1/2}$, or equivalently that
the characteristic Rossby number is approximately proportional to
$(Ra_F^*/E)^{1/2}$.   Increases or decreases in $Ra_F^*$ and $E$ 
by the same factor leave $Ro$ unchanged; in other words, this 
theory predicts that
constant-$Ro$ contours should have a slope of one in the $Ra_F^*$--$E$
plane.  This explains the fact that the empirically determined
slopes in the simulations are close to one (see Fig.~\ref{rossby-2D-sims}).
In our theory, this behavior occurs because the forcing depends
on the flux in the same way as damping depends on viscosity
(namely, linearly).  Thus, equal increases (or decreases) in the flux 
and the viscosity cause alterations to the forcing and damping
that cancel out, leading to a mean jet speed that is unchanged.

To further compare the theory to the simulations, 
Fig.~\ref{rossby-2D-viscous} plots our 
solution (Eq.~\ref{rossby-viscous}) as contours of Rossby number 
versus $Ra_F^*$ and $E$.  A choice of $k D$ is necessary to 
evaluate Eq.~(\ref{rossby-viscous}).  \citet{kaspi-etal-2009}'s simulations 
show that, although the jets themselves
are broad (with widths comparable to the domain thickness), 
the {\it curvature} of the wind $\nabla^2 u$, and thus the
dissipation, exhibits a dominant wavelength of about $D/5$
(see their Fig.~10d).  High-wavenumber structure in the flow
shear also occurs in \citet{christensen-2002}'s simulations.  To account
for this high-wavenumber structure, we adopt  $kD\approx 5\pi$,
and we calculate Fig.~\ref{rossby-2D-viscous} using this value
and $\epsilon=0.3$. 

 Figure~\ref{rossby-2D-viscous} reiterates
the approximate agreements in the slopes of constant-$Ro$
contours (compare Fig.~\ref{rossby-2D-viscous} with 
Fig.~\ref{rossby-2D-sims}).  Moreover,
the $Ro$ values themselves also agree well, particularly
in the regime of fast zonal jets where $Ro\gtrsim 0.02$.
Throughout this part of the parameter space, 
the predicted Rossby number (Fig.~\ref{rossby-2D-viscous}) matches 
those obtained in the simulations (Fig.~\ref{rossby-2D-sims}) 
to within a factor of $\sim$2 at a given $Ra_F^*$ and $E$,
although the discrepancy approaches an order of magnitude
toward the left side of the plot (at $Ro \lesssim 0.02$)
where the simulated Rossby numbers seem to undergo a regime shift
that is not captured by Eq.~(\ref{rossby-viscous}).
Such a shift could result for example from a dependence of $\epsilon$ 
or $k$ on $Ra_F^*$ and $E$, a possibility not considered up to now.   We
return to this point in the next two subsections.

\subsection{Regime II: Weakly nonlinear regime}
\label{momentum-approach}

For sufficiently small values of $Ra_F^*$, both the Boussinesq
and anelastic simulations shown in Fig.~\ref{rossby-2D-sims}
exhibit a regime shift where the dependence 
of Rossby number on $Ra_F^*$ and $E$ becomes steeper than the 
$(Ra_F^*/E)^{1/2}$ dependence discussed in the previous subsection.
As Fig.~\ref{rossby-2D-sims} shows, at Rossby numbers smaller than
$\sim$0.01--0.02, the constant-$Ro$ contours becomes closely spaced,
with a spacing indicating that a scaling $(Ra_F^*/E)^\xi$ with 
$\xi\approx1$ might provide an approximate fit.   
There is also an indication that the slopes of the constant-$Ro$ 
contours decrease toward the left side of the plot, particularly
for Christensen's simulations.

These stronger dependences of Rossby number on $Ra_F^*$
suggest a regime shift where different
processes set the characteristic Rossby number at low Rossby number
than at high Rossby number.  In the strongly nonlinear
regime explored in the previous subsection, the Rayleigh
number is strongly supercritical, and the convection is chaotic and 
nonlinear.   At lower Rayleigh numbers, when the Rayleigh number
is only modestly greater than the critical value for convection,
the convection is laminar and more spatially organized.  Here,
we explore how this transition in convective behavior
might lead to the regime shift in the jet
speeds seen in Fig.~\ref{rossby-2D-sims}.

To do so, we consider an alternate approach based the zonal momentum 
balance.  Consider a cylindrical
coordinate system whose axis aligns with the rotation axis,
with coordinates $(s, \lambda, z)$ corresponding to cylindrical radius
(i.e., distance from the rotation axis), longitude, and distance above or
below the equatorial plane, respectively.  The zonal-mean zonal
momentum equation then reads \citep[cf][]{kaspi-etal-2009}
\begin{eqnarray}
{\partial\overline{u}\over\partial t} + 2\Omega\overline{v_s} + 
{1\over \tilde{\rho}}\nabla\cdot(\tilde{\rho}\,\overline{u}\,
\overline{\bf v}) + {1\over \tilde{\rho}}\nabla\cdot
(\overline{\tilde{\rho}\, u' {\bf v'}})
=\nu \nabla^2 \overline{u}
\label{zonal-momentum}
\end{eqnarray}
where $u$ is the zonal wind, $v_s$ is the cylindrically radial wind
component (that is, the velocity component away from the rotation axis),
overbars denote zonal means, primes deviations therefrom, and
$\tilde{\rho}$
is the mean density (constant in the Boussinesq case and a specified
function of radius in the anelastic case).  On the left side, the second
term is the Coriolis acceleration.  The third term represents 
advection due to the mean-meridional flow, and the fourth term
represents the acceleration caused by eddies (i.e., Reynolds stress
convergences).  Generally speaking, for the
geostrophic jovian regime, the Reynolds stress term dominates
over the advection term \citep[e.g.][]{kaspi-etal-2009}. 

We now average the equation in $z$ (i.e. along the direction of
Taylor columns). The Coriolis accelerations cancel out, because
mass continuity prohibits any net, column-averaged motion toward or
away from the rotation axis.  Coupled with the free-slip boundary
conditions and symmetry of the zonal-wind structure about the equatorial 
plane, the $z$-averaging also removes the $z$-components of the 
divergence and Laplacian terms.  Neglecting the mean-flow 
advection terms, the column-averaged, steady-state
balance becomes
\begin{equation}
{1\over s^2 \tilde{\rho}}{\partial\over \partial s}
(s^2 \overline{u' v_s' \tilde{\rho}})
\approx \nu {\partial\over\partial s}\left[ {1\over s}{\partial\over\partial s}
(s \overline{u})\right]
\label{zonal-momentum2}
\end{equation}
which states that, in steady state, viscous drag associated with
shear of the mean zonal wind $\partial \overline{u}/\partial s$ balances
jet acceleration caused by eddies.  Balances analogous to this expression have
been written, for example, by \citet{busse-hood-1982}, \citet{busse-1983}, 
\citet{busse-1983c}, and 
\citet{cardin-olson-1994}.\footnote{An analogous balance was considered
by \citet{aubert-etal-2001} and \citet{aubert-2005} but for the no-slip
boundary condition where boundary-layer friction
plays the dominant role in the damping.}

We here use this balance to achieve a simple expression for the jet
speeds.  Approximate the left-hand side as $k\,\overline{u'v_s'}$, where
$k$ is the wavenumber (in cylindrical radius) over which $\overline{u'v_s'}$
varies significantly.  Furthermore, suppose that {\it both}
$u'$ and $v_s'$ scale like our expected convective velocity scale,
so that $\overline{u' v_s'}\sim C w^2$, where $w$ is the convective
velocity (e.g., from Eq.~\ref{vert-velocity}) and $C$ is a correlation
coefficient equal to one when $u'$ and $v_s'$ are perfectly
correlated (i.e. when eastward $u'$ always occurs with outward $v_s'$
and vice versa) and equal to zero when $u'$ and $v_s'$ exhibit
no correlation.  We approximate the right-hand side
as $\nu k^2 u$.  The momentum balance then becomes
\begin{equation}
u \sim C {w^2\over \nu k}
\label{u-reynolds}
\end{equation}
Nondimensionalizing, the Rossby number associated with the
jets is
\begin{equation}
Ro \sim {C \over kD} {Ro_{\rm conv}^2\over E}
\label{rossby-reynolds}
\end{equation}
An equation analogous to this was derived by \citet[his Eq.~3.12]
{christensen-2002}.
Inserting our expression for the convective velocities 
(Eq.~\ref{vert-velocity-nondim}) yields
\begin{equation}
Ro \sim {C \gamma^2 \over kD} {Ra_F^{\Delta}\over E}.
\label{rossby-reynolds3}
\end{equation}

Thus, this theory predicts that, if the degree of correlation 
between $u'$ and $v_s'$ is independent of the control parameters 
(i.e., if $C$ is constant), then the domain-mean jet speed scales 
as the convective flux over the viscosity, 
or equivalently the mean Rossby number 
scales as  $(Ra_F^*-Ra_F^{*\rm crit})/E$.  Away from the 
critical Rayleigh number, increases or decreases in $Ra_F^*$
and $E$ by the same factor leave $Ro$ unchanged; therefore,
as with the theory presented in the previous subsection, 
Eq.~(\ref{rossby-reynolds3}) predicts that constant-$Ro$ contours
should have slopes close to one in the logarithmic $Ra_F^*$--$E$ plane.
Near the critical Rayleigh number, however, the predicted slopes
differ from one.  If the critical modified flux Rayleigh number
depends on $E$ to a positive power less than one, then the constant-$Ro$
contours increase in slope, becoming more vertical; if, however, 
$Ra_F^{*\rm crit}$ depends on $E$ to a power greater than one, 
then the constant-$Ro$ contours decrease in slope, becoming more horizontal.

To evaluate Eq.~(\ref{rossby-reynolds3}) quantitatively, 
we require an expression for $Ra_F^{*\rm crit}$.  By performing many 
simulations, \citet{christensen-2002} determined empirically the 
critical Rayleigh number for each Ekman number he explored.\footnote{
Christensen found that the critical values of a modified Rayleigh number
$Ra^*\equiv Ra E^2 Pr^{-1}$ are 0.001005, 0.002413, 0.006510,
and 0.016790 for Ekman numbers of $10^{-5}$, $3\times10^{-5}$,
$10^{-4}$, and $3\times10^{-4}$, respectively.  This is equivalent
to modified flux critical Rayleigh numbers, $Ra_F^{*\rm crit}$, 
of $1.005 \times 10^{-8}$, $7.2\times10^{-8}$, $6.5\times10^{-7}$,
and $5.04\times10^{-6}$ for those same Ekman numbers, respectively.
Equation~(\ref{Racrit}) gives values that agree with these to within
$\sim$10\%.}
\citet{cardin-olson-1994} performed a linear instability analysis
of rotating convection in a spherical shell and found that,
at $E\ll1$, the
critical value of the ordinary Rayleigh number is a constant 
times $E^{-17/15} Pr^{4/3} (1+Pr)^{-4/3}$.  
Using the relationships between $Ra$ and $Ra_F^*$ 
(see Section~\ref{overforcing}), and noting that the Nusselt number
equals one at the onset of convection, shows that the critical modified
flux Rayleigh number should then scale as $E^{28/15} Pr^{-2/3} (1+Pr)^{-4/3}$.
Most of Christensen's simulations adopt a Prandtl number of one and
we find that his empirically determined critical Rayleigh numbers are
well matched by the expression\footnote{\tt \citet{dormy-etal-2004} found
theoretically that the critical Rayleigh number should scale as $E^{-4/3}$
at low Ekman number, which would imply that $Ra_F^{*\rm crit}$ should scale
as $E^{5/3}$.  However, this scaling does not fit Christensen's empirically
determined critical Rayleigh numbers as well as Eq.~(\ref{Racrit}), so we
retain scaling (\ref{Racrit}) for the purposes of this paper.  The
choice has a negligible influence on our results.}
\begin{equation}
Ra_F^{*\rm crit} \approx 20 E^{28/15}.
\label{Racrit}
\end{equation}
Given this dependence, contours of constant Rossby number in the
logarithmic $Ra_F^*$--$E$ plane should
decrease in slope as one approaches the critical Rayleigh
number (i.e., as one moves toward lower $Ra_F^*$ values).  This 
is qualitatively consistent with the behavior exhibited by
Christensen's simulations (see Fig.~\ref{rossby-2D-sims}).

\subsection{Combining Regimes I and II}

We have discussed two distinct regimes: (i) a regime of fast zonal
winds and strongly supercritical convection where the simulated
jet speeds seem well explained by the assumption that $\epsilon$
is constant (Regime~I), and (ii) a regime of slow zonal 
winds and weakly supercritical convection where the jet speeds
are approximately explained by the assumption that the correlation
between the $u'$ and $v_s'$ velocity components is strong and 
roughly constant (Regime~II).  These led to two distinct
scalings for the jet speeds: Eqs.~(\ref{rossby-viscous}) and
(\ref{rossby-reynolds3}) for the two regimes, 
respectively.  Two issues now arise:
First, how do we combine these regimes?  Second, why should
energetics and momentum considerations lead to different scalings?
Both approaches should, in principle, yield the same answer within
any given regime.

\begin{figure}
\includegraphics[scale=0.48]{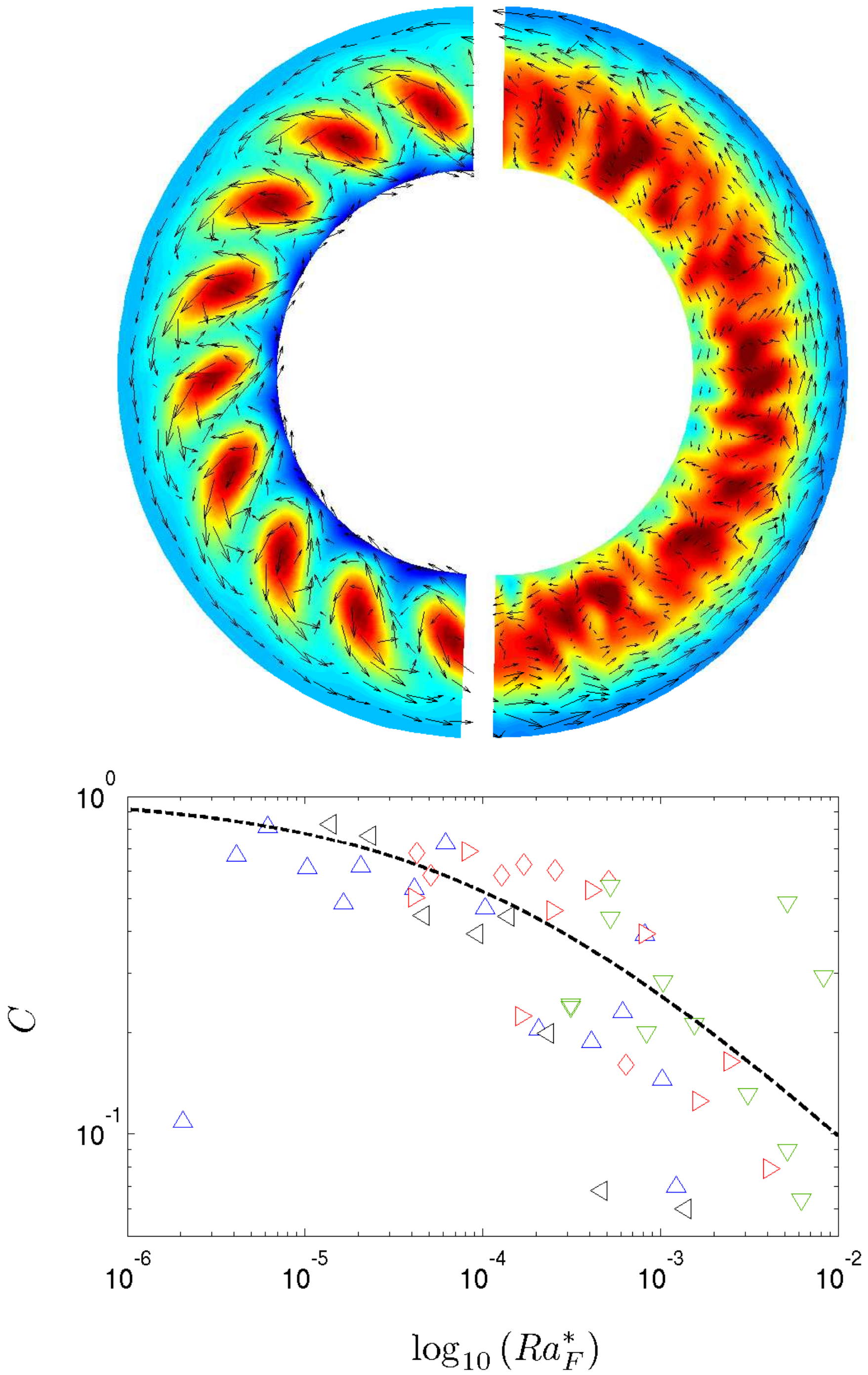}
\caption{{\it Top:} Snapshots in the equatorial plane of two anelastic, 3D 
simulations from \citet{kaspi-2008} illustrating how the correlation
between the convective velocity components depends on the supercriticality.
Each slice shows half of the equatorial plane of one simulation 
(note however that the simulations each span $360^{\circ}$ of longitude).
Color depicts streamfunction and arrows denote velocity component in the
equatorial plane.  The left simulation is weakly supercritical; the
strong correlation between outward and eastward velocity components
is obvious.  The right simulation is strongly supercritical; the
correlation between the outward and eastward velocity components is 
weaker because of the complex, turbulent convective structure.
{\it Bottom:} Depicts correlation coefficient $C$ versus $Ra_F^*$
for a range of anelastic simulations$^{\ref{eps-c-footnote}}$.  
As expected qualitatively from the top
panels, $C$ decreases with $Ra_F^*$.  Symbols are defined
in Table~\ref{symbol-key}.}
\label{weakly-strongly}
\end{figure}

The resolution to both issues lies in the 
dependence of the correlation coefficient $C$ and the 
convective jet-pumping efficiency $\epsilon$ on the control 
parameters.  Consider, for example, the correlation coefficient.
When the Rayleigh number is weakly supercritical and the flow is 
geostrophic, the convection forms broad, laminar convection 
rolls, which in a sphere become tilted in the prograde direction
\citep[and others]{busse-hood-1982, busse-1983, busse-1983c,
busse-2002, tilgner-busse-1997, sun-etal-1993, christensen-2002, 
kaspi-2008}.  This leads to highly correlated velocity 
components $u'$ and $v_s'$, 
as illustrated in Fig.~\ref{weakly-strongly} ({\it top left}):
ascending fluid parcels move prograde 
while descending parcels move retrograde.  As a result, at weakly
supercritical Rayleigh numbers, the correlation
coefficient $C$ should be close to one, at least outside the
tangent cylinder.  On the other hand, when the convection is
strongly supercritical, the convective structure is chaotically
time-dependent and spatially complex; 
the velocity components $u'$ and $v_s'$ are
positively correlated in some regions but negatively correlated
in others.  This is illustrated in Fig.~\ref{weakly-strongly}
({\it top right}). The net correlation is usually still positive, but
the large degree of cancellation implies that, at strongly
supercritical Rayleigh numbers, the correlation coefficient 
should drop significantly below one \citep{christensen-2002}.
As a result, one expects the correlation coefficient
$C$ to vary slowly with $Ra_F^*$ near the critical Rayleigh
number but to decrease more rapidly with increasing $Ra_F^*$ at sufficiently
supercritical Rayleigh numbers.   This behavior can be
seen in correlation coefficients calculated for a range of
simulations versus $Ra_F^*$ (Fig.~\ref{weakly-strongly}, {\it bottom}).

Next consider the jet-pumping efficiency $\epsilon$.  The scaling
(\ref{rossby-viscous}), wherein the Rossby number scales with
$(Ra_F^*/E)^{1/2}$, assumes that $\epsilon$ is constant.  While
this assumption seems to work well at sufficiently supercritical
Rayleigh numbers, it {\it must} fail in the weakly supercritical
regime where $u'$ and $v_s'$ are highly correlated. This is because,
in this weakly nonlinear regime, the mean jet speeds scale
quadratically with the convective velocities, and the
power exerted to pump the jets scales as the Reynolds stresses times
the jet speeds, namely as the fourth power of the convective velocities.
The dependence of this quantity on the heat flux differs from
the dependence of the convective potential-energy release on
the heat flux.  Since $\epsilon$ is the ratio of these quantities,
$\epsilon$ cannot be constant in this regime.

A regime shift analogous to that seen for the jet speeds
in Fig.~\ref{rossby-2D-sims}---where Rossby numbers scale approximately
with $Ra_F^*/E$ at low Rossby number and approximately as
$(Ra_F^*/E)^{1/2}$ high Rossby number---can occur if the correlation 
coefficient depends weakly on $Ra_F^*$ at low $Ra_F^*$ yet 
strongly on $Ra_F^*$ at high $Ra_F^*$.  Likewise, it can occur 
if the jet-pumping efficiency
depends strongly on $Ra_F^*$ at low $Ra_F^*$ yet weakly at high 
$Ra_F^*$.

To quantify these arguments, let us relate $\epsilon$ and $C$
to each other.  By definition, the efficiency $\epsilon$ is the ratio of work
done to pump the jets to the work made available by convection: 
\begin{equation}
\epsilon \sim 
{\overline{u}{1\over\tilde{\rho}}\nabla\cdot(\overline{\tilde{\rho}u'{\bf v'}})
\over gw\alpha \, \delta T},
\end{equation}
where it is understood that both the numerator and the denominator
represent global averages.  Expressing the numerator as $C k w^2$,
the denominator as $g\alpha F/\rho c_p$ (using Eq.~\ref{heat-transport}), 
and non-dimensionalizing, we obtain
\begin{equation}
\epsilon \sim {Ro C k D Ro_{\rm conv}^2\over Ra_F^{\Delta}}.
\label{epsilon}
\end{equation}
Using Eq.~(\ref{rossby-reynolds}) for the Rossby number yields
the expression
\begin{equation}
\epsilon \sim  C^2 {Ro_{\rm conv}^4\over E Ra_F^{\Delta}}.
\label{epsilon2}
\end{equation}
If we adopt $Ro_{\rm conv}\approx \gamma(Ra_F^{\Delta})^{1/2}$ for
concreteness, we obtain finally
\begin{equation}
\epsilon \sim C^2 \gamma^4 {Ra_F^{\Delta}\over E}.
\label{epsilon3}
\end{equation}
Therefore, $\epsilon$ and $C$ cannot simultaneously be constant when
$Ra_F^*$ or $E$ are varied.  Holding one constant requires the
other to become a function of $Ra_F^*$ and $E$.

Let us now make the 
simplest possible assumption regarding this regime shift---we postulate
a regime with $C\approx \,{\rm constant}=1$ at low Rayleigh number
and a regime with $\epsilon\approx \,{\rm constant}\equiv\epsilon_{\rm max}$ at
high Rayleigh number. Quantitatively, there is no rigorous
expectation that $C$ or $\epsilon$ need be constant, nor (as
described previously) is this assumption actually necessary for 
a regime shift to occur. 
Nevertheless, $C$ and $\epsilon$ have upper limits of 1, so if
some process causes them to increase with increasing (or decreasing)
$Ra_F^*$, they might naturally plateau---at least over some range
of parameter space---upon approaching their upper limits.
Still, future theoretical work on
what sets the $Ra_F^*$- and $E$-dependence of the correlation
coefficient and jet-pumping efficiency is warranted.

Given these postulates, Eq.~(\ref{epsilon3}) implies that
\begin{equation}
C = \cases{1, &$Ra_F^{\Delta} \lesssim \epsilon_{\rm max}\gamma^{-4} E$;\cr 
\epsilon_{\rm max}^{1/2} \gamma^{-2}\left({E\over Ra_F^{\Delta}}\right)^{1/2},&$Ra_F^{\Delta} \gtrsim \epsilon_{\rm max}\gamma^{-4}E$\cr} 
\label{corr-coeff-cases}
\end{equation}
and
\begin{equation}
\epsilon = \cases{\gamma^4 {Ra_F^{\Delta}\over E}, &$Ra_F^{\Delta} \lesssim \epsilon_{\rm max}\gamma^{-4}E$;\cr 
\epsilon_{\rm max},&$Ra_F^{\Delta} \gtrsim \epsilon_{\rm max}\gamma^{-4}E$\cr} 
\label{epsilon-cases}
\end{equation}
To smoothly transition between the regimes, we adopt the 
expression
\begin{equation}
C \sim \left[1 + {\gamma^2\over\epsilon_{\rm max}^{1/2}}\left({Ra_F^{\Delta}\over
E}\right)^{1/2}\right]^{-1}
\label{corr-coeff-smooth}
\end{equation}
which is merely the inverse of one over the value of $C$ from the
first regime plus one over the value of $C$ from the second regime.
This gives behavior equivalent to Eq.~(\ref{corr-coeff-cases}) in
the two limits with a smooth transition in between.  Given this
expression, $\epsilon$ can then be evaluated from Eq.~(\ref{epsilon3}),
yielding
\begin{equation}
\epsilon \sim \gamma^4
\left[1 + {\gamma^2\over\epsilon_{\rm max}^{1/2}}\left({Ra_F^{\Delta}\over
E}\right)^{1/2}\right]^{-2}{Ra_F^{\Delta}\over E}.
\label{epsilon-smooth}
\end{equation}

\begin{figure}
\includegraphics[scale=0.45]{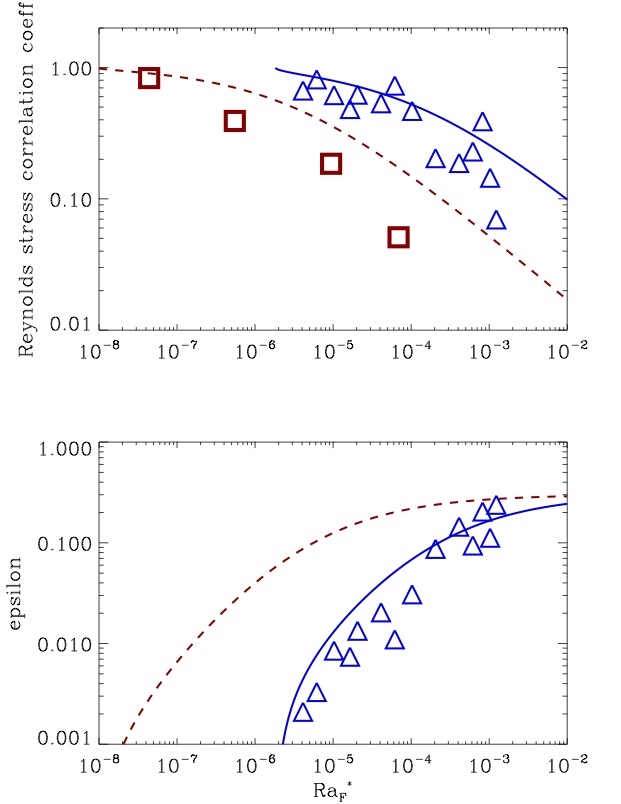}
\caption{{\it Top:} The correlation coefficient $C$, the ratio
between the Reynolds stress
$\overline{u'v_s'}$ and the product of the root-mean-square velocity
magnitudes.  {\it Bottom:} $\epsilon$, the fraction of potential 
energy released by convection that is used to pump the jets. 
Blue triangles depict values we calculated rigorously from 
anelastic simulations at $E=4\times10^{-4}$, while red squares
denote values \citet{christensen-2002} calculated rigorously
from Boussinesq simulations at $E=10^{-5}$.  In both cases, values
represent flow behavior outside the tangent cylinder.  Curves plot
Eqs.~(\ref{corr-coeff-smooth})--(\ref{epsilon-smooth}) at 
an Ekman number of $4\times10^{-4}$ (solid blue) and $10^{-5}$
(dashed red) assuming $\gamma=1$ and $\epsilon_{\rm max}=0.3$,
with $Ra_F^{*\rm crit}$ evaluated using $20E^{28/15}$ for the Boussinesq
simulations and $0.4E^{28/15}$ for the anelastic simulations (the
difference resulting from the differing Prandtl numbers; see 
Table~\ref{symbol-key}). }
\label{corr-coeff-epsilon}
\end{figure}

Figure~\ref{corr-coeff-epsilon} plots $\epsilon$ and $C$
versus $Ra_F^*$ from available data and compares them to 
Eqs.~(\ref{corr-coeff-smooth})--(\ref{epsilon-smooth}).
In particular, the blue triangles denote values of $C$ and
$\epsilon$ that we rigorously computed\footnote{\label{eps-c-footnote}
The correlation
coefficient and jet-pumping efficiency were calculated from 
the anelastic simulations as 
\begin{eqnarray}\nonumber
C =  {[\overline{u'v_s'}]\over [(\overline{u'^2\,v'^2})^{1/2}]}
\end{eqnarray}
and
\begin{eqnarray}\nonumber
\epsilon = 
{[\overline{u}{1\over\tilde{\rho}}\nabla\cdot(\overline{\tilde{\rho}u'
{\bf v'}})]
\over [\overline{gw\alpha \, \delta T}]}
\end{eqnarray}
where $[...]=\int(...)\tilde{\rho}s\,ds/\int \rho s ds$ indicates
a mass-weighted average over the equatorial plane.} from the
three-dimensional convective velocity and entropy fields
for a sequence of anelastic simulations with 
$E=4\times10^{-4}$.
  Additionally, the red circles denote values of $C$ rigorously 
calculated by \citet{christensen-2002} for 
a sequence of his simulations at $E=10^{-5}$ (see his Table 3; note
that no data on $\epsilon$ are available for Christensen's simulations.)  
Blue and red solid curves denote the predictions of 
Eqs.~(\ref{corr-coeff-smooth})--(\ref{epsilon-smooth}) for $E=4\times10^{-4}$
and $10^{-5}$, respectively, and can be directly compared to the
symbols of the same color.

Key points are as follows.  First, the data indicate that,
for both sets of simulations, $C$ decreases with $Ra_F^*$, while 
$\epsilon$ increases with $Ra_F^*$.  Second, and more 
importantly, the {\it slopes}
of the trends change across the range of $Ra_F^*$ explored.  The trends 
of $C$ in both sets of simulations depend relatively weakly on $Ra_F^*$ 
toward the left but seem to steepen toward the right.  For
example, Christensen's data indicate that a factor-of-two change
in $Ra_F^*$ cause a 1.28-fold change in $C$ at low $Ra_F^*$
but a 2.7-fold change in $C$ at high $Ra_F^*$.  Conversely,
the data suggest that $\epsilon$ depends on $Ra_F^*$ strongly at 
small $Ra_F^*$ but more weakly at larger $Ra_F^*$.  As Eq.~(\ref{epsilon3})
shows, these trends must go hand-in-hand; consistency between
the two quantities requires the trend in $C$ to steepen if that
in $\epsilon$ flattens.  Moreover, for both $C$ and 
$\epsilon$, these changes in the slopes with $Ra_F^*$
are precisely the ingredients
that can generate a regime shift wherein Rossby numbers depend strongly
on $Ra_F^*$ at low $Ra_F^*$ (as postulated for Regime II in 
Section~\ref{viscous}{\it b}) but more weakly on $Ra_F^*$ at high
$Ra_F^*$ (as postulated for Regime I in Section~\ref{viscous}{\it a}).
Third, the data suggest that, at constant $Ra_F^*$, the correlation
coefficient increases with increasing $E$.

The theoretical curves for $C$ and $\epsilon$ match the data surprisingly 
well (Fig.~\ref{corr-coeff-epsilon}).  
All the qualitative trends discussed above
for the data are predicted by the theory: $C$ decreases and $\epsilon$
increases with $Ra_F^*$; the slope of $C(Ra_F^*)$ steepens toward the
right while that of $\epsilon(Ra_F^*)$ flattens toward the right;
and at constant $Ra_F^*$ the values of $C$ increase with $E$.
The theory also predicts that at constant $Ra_F^*$, the values of $\epsilon$
should decrease with increasing $E$, which can be tested with future
simulations.  Moreover, for the parameters
chosen ($\gamma=1$ and $\epsilon_{\rm max}=0.3$), the numerical values 
of the theoretical $\epsilon$ and $C$ curves agree relatively well
with the simulation data, particularly for the anelastic simulations
at $E=4\times10^{-4}$.  On the other hand, the theoretical correlation 
coefficients at $E=10^{-5}$ overpredict those for the corresponding Boussinesq
simulations by up to a factor of several, particularly at the higher-$Ra_F^*$
values.  Of course, the technique adopted to smooth between the two regimes
(cf Eqs.~\ref{corr-coeff-smooth}--\ref{epsilon-smooth}) will affect
the degree to which the curves agree with the data, so the details
of the comparisons should be considered tentative.

\begin{figure}
\includegraphics[scale=0.4]{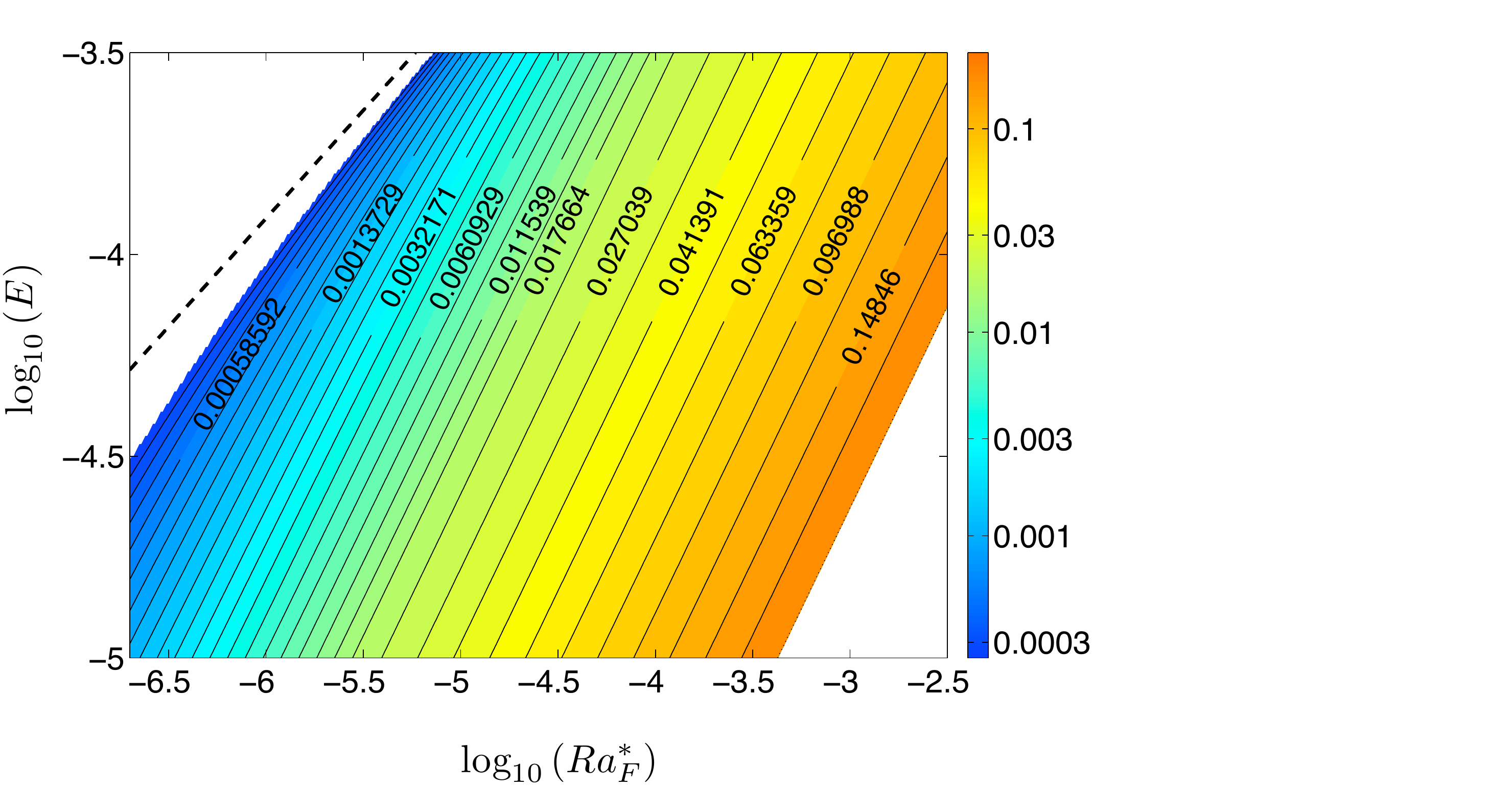}
\caption{Prediction of our scaling law, 
Eq.~(\ref{two-viscous-regimes-equation}), that combines Regimes I and
II.  Depicts contours of Rossby number versus modified flux Rayleigh
number (abscisssa) and Ekman number (ordinate).  Adopts $kD=5\pi$,
$\gamma=1$, and
$\epsilon_{\rm max}=0.3$ (see text).  Note the similarity of the slopes and
values of the contours with the simulation results plotted in 
Fig.~\ref{rossby-2D-sims}.  Contours are evenly
spaced in $\log(Ro)$, with contour values identical to those in
Figs.~\ref{rossby-2D-sims} and \ref{rossby-2D-viscous}.  For reference, 
the dashed line depicts the
critical modified flux Rayleigh number given by Eq.~(\ref{Racrit}).
Above the dashed line, convection does not occur.}
\label{two-viscous-regimes}
\end{figure}

What are the implications of this discussion for the jet speeds?
Inserting our expression for $\epsilon$ into Eq.~(\ref{rossby-viscous})
or our expression for $C$ into Eq.~(\ref{rossby-reynolds3}) 
leads to a single, self-consistent expression for the
Rossby number spanning both regimes.  For simplicity, 
we adopt Eqs.~(\ref{corr-coeff-cases})--(\ref{epsilon-cases}) rather
than their smoothed counterparts, leading to
the expression
\begin{equation}
Ro = {1\over kD} \min\left[ \gamma^2{Ra_F^* - Ra_F^{*\rm crit}\over E}, 
\,\epsilon_{\rm max}^{1/2}
\left({Ra_F^* - Ra_F^{*\rm crit}\over E}\right)^{1/2}\right].
\label{two-viscous-regimes-equation}
\end{equation}
which is equivalent to the minimum of Eqs.~(\ref{rossby-viscous})
and (\ref{rossby-reynolds3}).
Thus, arguments based on energetics and momentum arguments are 
now self-consistent: both approaches predict that the Rossby number
scales as $(Ra_F^* - Ra_F^{*\rm crit})/E$ at weakly supercritical
Rayleigh numbers but as $(Ra_F^*-Ra_F^{*\rm crit}/E)^{1/2}$ at strongly supercritical
Rayleigh numbers.

Figure~\ref{two-viscous-regimes} plots the jet speed predicted
by Eq.~(\ref{two-viscous-regimes-equation}) as contours of
Rossby number versus $Ra_F^*$ and $E$, with a contour spacing
that is even in $\log(Ro)$ and contour values identical to those
in Figs.~\ref{rossby-2D-sims} and \ref{rossby-2D-viscous}. 
Regime I, in which the jet-pumping efficiency $\epsilon$
is constant, lies toward the lower right, while Regime II, in
which the correlation coefficient $C$ is constant, lies toward the
upper left.  The transition between the regimes manifests visually
as a jump in the spacing of the constant-$Ro$ contours and occurs 
at a Rossby number of $\sim$0.02 for the parameters chosen, 
roughly consistent with the regime transition in the simulated data
(see Fig.~\ref{rossby-2D-sims}).
Within Regime II, the slopes of the contours decrease 
as the Rossby number is decreased and the modified flux Rayleigh
number approaches the critical value.   A similar decrease in the slope
of the contours at small Rossby number is evident in \citet{christensen-2002}'s
data plotted in Fig.~\ref{rossby-2D-sims}.  Moreover, throughout
most of the parameter space accessed by the simulations, the 
predicted Rossby number (Fig.~\ref{two-viscous-regimes}) matches
those obtained in the simulations (Fig.~\ref{rossby-2D-sims}) within
a factor of $\sim$2.

\section{Scaling for the jet speeds in Regime III: the asymptotic regime}
\label{asymptotic}

The simulated behavior in Fig.~\ref{rossby-2D-sims} is reasonably
well explained by the theory presented in Section~\ref{viscous},
where Rossby numbers scale approximately with  $(Ra_F^*-Ra_F^{*\rm crit})/E$ 
at low Rayleigh numbers (Regime II) and as $(Ra_F^*/E)^{1/2}$ at high
Rayleigh numbers (Regime I). However, \citet{christensen-2002} 
argued that at the highest $Ra_F^*$ he explored for each value of $E$
(or equivalently the lowest $E$ values explored for each value 
of $Ra_F^*$) his simulations approached an asymptotic regime where
the mean equilibrated jet speed became independent of the values of 
the diffusivities and depend only weakly on the heat flux.
This behavior manifests in Fig.~\ref{rossby-2D-sims}({\it top}) 
as both a widening of the constant-$Ro$ contours and significant
increase in their slope beyond one toward the far right edge of
the plot, neither of which is predicted by the scalings in
Section~\ref{viscous}.  Based on an empirical fit 
to his simulation results, \citet{christensen-2002} proposed that 
in this asymptotic limit---which we call Regime III---the Rossby 
number associated with the characteristic jet speeds scales 
as\footnote{Because he 
included an inner-to-outer radius ratio in his definition of 
$Ra_F^*$, \citet{christensen-2002} quotes the relationship with a 
pre-factor of 0.65 rather than 0.53.}
\begin{equation}
Ro = 0.53(Ra_F^*)^{1/5}.
\label{christensen}
\end{equation}

At present, however, a physical basis for this empirical fit 
is lacking.  To see whether Eq.~(\ref{christensen}) can be
theoretically explained, we present here a scaling analysis based on 
mixing-length theory.  We compare it to available simulations and
evaluate its implications for the Jovian parameter regime.  {\tt
Nevertheless, we emphasize at the outset that the existence of
the asymptotic regime is tentative, and further numerical work is
required to confirm (or refute) its existence and determine its properties.
This section is intended simply to offer some ideas on how such
an asymptotic regime, if it exists, could work.}


We hypothesize that, if the viscosity is sufficiently small,
the damping will be determined not by the molecular viscosity but
by an eddy viscosity associated
with turbulent diffusion.  The concept of an 
eddy viscosity is relevant only when the convection is
sufficiently supercritical (so that the eddies behave in a nonlinear,
turbulent manner), leading us to adopt the approach of Section 4{\it a}.
To obtain an expression for mean wind speed, 
we again equate forcing (Eq.~\ref{integrated-forcing2}) and 
damping (Eq.~\ref{integrated-loss}) but use the eddy viscosity,
$\nu_{\rm eddy}$, in place of $\nu$ in Eq.~(\ref{integrated-loss}).
We adopt a standard mixing-length formulation for the eddy
viscosity, $\nu_{\rm eddy}\approx w k_{\rm mix}^{-1}$, where $w$ 
is the convective velocity (given by 
Eqs.~\ref{vert-velocity}--\ref{vert-velocity-nondim})
and $k_{\rm mix}$ is the wavenumber associated with
the mixing length (e.g., a typical distance 
traversed by coherent convective plumes).  In the case of
a Boussinesq fluid, where $\alpha$ 
and $\rho$ are constant and $F$, $g$, and $c_p$ are approximately
constant, this yields
\begin{equation}
u\sim{\epsilon^{1/2}\Omega^{1/4} k_{\rm mix}^{1/2}\over k \gamma^{1/2}}
\left({\alpha g F\over \rho c_p}\right)^{1/4}
\label{u-asymp-boussinesq}
\end{equation}
where $\epsilon$ is assumed approximately constant as in Section 4{\it a}.
For an anelastic fluid, where $\alpha$ and $\rho$ vary greatly
with radius, we follow the approach used in Section~\ref{viscous}{\it a},
yielding
\begin{equation}
u \sim {{\epsilon}^{1/2} \Omega^{1/4} \over k \gamma^{1/2}} 
\left[{\int{\alpha g F\over c_p}r^2\,dr \over 
\int \left({\alpha g F\over c_p}\right)^{1/2} k_{\rm mix}^{-1} \rho^{1/2} r^2\,dr}
\right]^{1/2}.
\label{u-asymp-anelastic}
\end{equation}

Let us nondimensionalize these expressions so that they can be
compared with existing Boussinesq and anelastic numerical results
and the empirical asymptotic scaling (Eq.~\ref{christensen}) 
proposed by \citet{christensen-2002}.
Defining the Rossby number as $Ro\equiv u/(\Omega D)$ and assuming
that $k_{\rm mix}$ is constant and that $Ra_F^*\gg Ra_F^{*\rm crit}$, 
Eqs.~(\ref{u-asymp-boussinesq})--(\ref{u-asymp-anelastic}) can be 
nondimensionalized to yield
\begin{equation}
Ro = {\epsilon^{1/2} \Gamma  (k_{\rm mix}D)^{1/2}\over kD \gamma^{1/2}} 
(Ra_F^*)^{1/4}
\label{rossby-asymp}
\end{equation}
where $Ra_F^*$ is given by Eq.~(\ref{Ra_Fstar}) for the Boussinesq 
case and Eq.~(\ref{Ra_Fstar-integrated}) for the anelastic case.
$\Gamma$ is a dimensionless constant 
that depends on the planet's basic-state radial density structure and
is given by
\begin{equation}
\Gamma = 
\left[{ (\int \rho r^2\,dr)^{1/2} \left(\int {\alpha g f\over c_p} r^2\,dr
\right)^{1/2} \over \int \left({\alpha g f\over c_p}\right)^{1/2} 
\rho^{1/2} r^2\,dr}\right]^{1/2}
\label{constant}
\end{equation}
where $f(r)$ is an order-unity dimensionless function that describes
the radial dependence of the heat flux, $F= F_0 f(r)$, and $F_0$
is a constant (independent of radius) that gives the characteristic
value of the heat flux (e.g., halfway through the layer). 
Note that, for the case of constant $\rho$, $\alpha$, $g$, flux, and
$c_p$, then $\Gamma\equiv1$.  Interestingly,
for the \citet{kaspi-etal-2009} model, $\Gamma=1.02$ despite the large
variation of $\alpha$ and $\rho$ with radius.\footnote{Given the 
integrals in the numerator and denominator of $\Gamma$, one can show that
$\Gamma$ tends to be close to 1 for a wide range of possible radial dependences of the
function $\alpha g f/c_p$---even when this function varies in radius by orders of
magnitude---as long as it has a smooth dependence, as occurs in the numerical
models and in Jupiter.}
 
\begin{figure}
\includegraphics[scale=0.4]{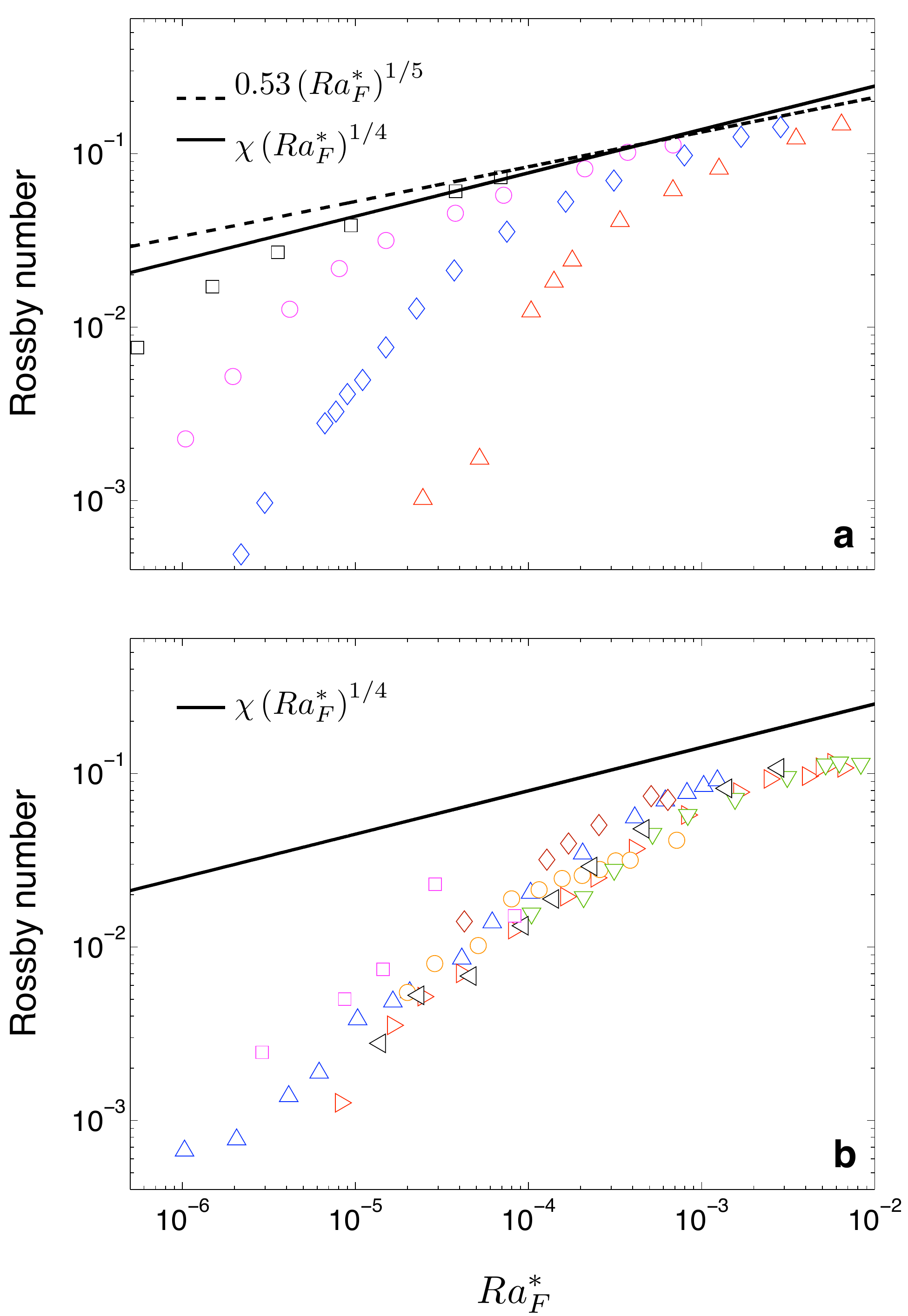}
\caption{Comparison of our asymptotic scaling for the jet speeds, 
Eq.~(\ref{rossby-asymp}) 
(solid curves) with Christensen's empirical asymptotic fit from
Eq.~(\ref{christensen}) (dashed curve) and simulation results from
\citet{christensen-2002} (panel {\it a}) and \citet{kaspi-etal-2009}
(panel {\it b}).  Solid curves are evaluated using 
$\epsilon=0.3$, $\gamma=1$, and $\chi=1$.  See Table~\ref{symbol-key}
for definitions of the symbols used in ({\it b}).}
\label{rossby-rayleigh}
\end{figure}

This simple theory therefore predicts that, in the asymptotic
regime, the characteristic jet speed scales as $F^{1/4}$, or
equivalently the characteristic Rossby number scales 
as $(Ra_F^*)^{1/4}$, with no explicit dependence on the molecular
or numerical viscosity.  Our exponent of 1/4 compares favorably to---but is
modestly steeper than---Christensen's empirically determined
exponent of 1/5.  Notably, our theoretical exponent is significantly
smaller than that for Regimes I and II (Section~\ref{viscous}),
where, at constant $E$, jet speeds scale with $F^{1/2}$ or $F$
depending on assumptions.
In our asymptotic theory, the weak dependence on $Ra_F^*$
occurs because the eddy viscosity decreases
with decreasing heat flux (unlike the behavior in Regimes I and II,
where the viscosity is a fixed parameter).  As a result, a weakly
forced flow has extremely weak damping, whereas a strongly forced
flow has extremely strong damping; this leads to an equilibrated jet
speed that depends only weakly on heat flux.  Specifically, the kinetic-energy
forcing scales with $F$, and because the convective velocities
scale with $F^{1/2}$ (Eq.~\ref{vert-velocity}), the eddy viscosity and therefore
the kinetic-energy damping also scale with $F^{1/2}$.  This leads to
an equilibrated kinetic energy scaling with $F^{1/2}$ and a jet
speed scaling with $F^{1/4}$.

Figure~\ref{rossby-rayleigh} explicitly compares the theoretical scaling
(Eq.~\ref{rossby-asymp}) with the empirical scaling (Eq.~\ref{christensen})
and the simulation results of \citet{christensen-2002} and
\citet{kaspi-etal-2009}.  The theoretical scaling is shown for
a value of the prefactor $\epsilon^{1/2} (k_{\rm mix}D)^{1/2} \gamma^{-1/2}
(kD)^{-1}$, which we call $\chi$,  
equal to 1.  For this value, the magnitudes of the jet speeds 
predicted by our Eq.~(\ref{rossby-asymp})
match Christensen's fit within a factor of $\sim$2
(compare solid and dashed lines in Fig.~\ref{rossby-rayleigh}a).
For the values of $kD\approx5\pi$ adopted previously, this would require
$k_{\rm mix}\approx (5\pi)^2/D$, implying a rather short mixing
length $2\pi k_{\rm mix}^{-1}$ of $\sim$$0.03D$.    This seems at 
least qualitatively consistent with the fact that the vorticity structure
of convective plumes exhibits little coherence in cylindrical radius 
$s$ \citep[see][]{christensen-2002}.  The eddy diffusivities implied 
by this choice of mixing length are similar to those suggested by previous authors, lending some
encouragement to this choice of prefactor.\footnote{The (dimensional) eddy diffusivity
implied by our expression for vertical velocity (Eq.~\ref{vert-velocity}) and the 
condition $\chi=1$ is $\nu_{\rm eddy}\approx
\epsilon D (\alpha g F)^{1/2}/[(\rho c_p\Omega)^{1/2} \gamma (kD)^2]$.
For Jovian interior conditions and heat flux ($\alpha\approx10^{-5}\rm\,K^{-1}$, 
$g\approx26\rm\,m\,sec^{-2}$, $F=5\rm\,W\,m^{-2}$, $\rho\approx1000\rm\,kg\,m^{-3}$,
$c_p\approx1.3\times10^4\rm\,J\,kg^{-1}\,K^{-1}$, $\Omega=1.74\times10^{-4}\rm\,sec^{-1}$
$D=2\times10^4\rm\,km$), and $\epsilon=0.3$ and $kD=5\pi$ as adopted
previously, this implies values of $\sim20\rm\,m^2\,sec^{-1}$.   This
is roughly consistent with the value of $\sim$$10\rm\,m^2\,sec^{-1}$ suggested
by \citet{starchenko-jones-2002}.}

Overall, the close agreement between the heat-flux dependences of 
our asymptotic scaling and Christensen's empirical fit
is encouraging and argues that the basic idea encapsulated by
our scaling---that the damping depends on $F$ to a power
modestly less than that of the forcing---is correct.

At high Rayleigh numbers, both the Boussinesq and anelastic
simulations (Fig.~\ref{rossby-rayleigh}a and b, respectively) converge
toward a trend similar to the shallow slope of the predicted 
asymptotic scaling.   However, the simulations diverge from
the asymptotic scalings at low Rayleigh numbers
as the effects of the model diffusivities become strong.  
Christensen's fit, and the scaling presented in this section, are
intended to describe the {\it asymptotic} behavior in the limit of 
negligible diffusivities \citep[for discussion see][]{christensen-2002}.

When extrapolated to Jovian values of $Ra_F^*$, this asymptotic regime 
predicts that the mass-weighted mean velocities in the molecular envelopes
of Jupiter and Saturn are small, $\sim$0.1--$1\rm\,m\,sec^{-1}$.  
We consider this issue further, and combine the three
regimes into a single scaling, in the next subsection.

\section{Combining the three scalings}
\label{combine}

We have derived scalings for the jet speeds in three regimes:
two regimes in which the numerical viscosity dominates the damping
(Regimes I and II in Sections~\ref{viscous}{\it a} and 
\ref{viscous}{\it b}, respectively)
and another in which the viscosity is determined by turbulence,
i.e., an eddy viscosity (Regime III in Section~\ref{asymptotic}).  
Here, we combine the scalings.

Our basic approach in deriving the scalings was to balance forcing 
against damping;
the damping represents a numerical (or molecular) viscosity in
Regime I but an eddy viscosity in Regime III.  Here, we suppose
that both molecular and eddy viscosities operate simultaneously,
and that whichever viscosity is larger will dominate.  Because
larger viscosity implies smaller equilibrated jet speeds, 
we therefore expect that  the {\it smaller} of the two Rossby numbers 
will dominate.  Combining Regimes I, II, and III, we can thus roughly say
\begin{eqnarray}\label{three-regimes} \nonumber
Ro = {1\over kD}\min\Biggl[\gamma^2{Ra_F^* - Ra_F^{*\rm crit}\over E}, \\
\epsilon_{\rm max}^{1/2} \left({Ra_F^*-Ra_F^{*\rm crit}\over E}\right)^{1/2}, 
\epsilon_{\rm max}^{1/2} \Gamma {(k_{\rm mix}D)^{1/2}\over\gamma^{1/2}} 
(Ra_F^*)^{1/4}\Biggr]
\end{eqnarray}
where we have assumed that $k$ is constant and $\epsilon_{\rm max}$
has the same values in the second and third terms on the right side.
Neglecting $Ra_F^{*\rm crit}$ in the expression for Regime I,
the transition between Regimes I and III occurs for Ekman numbers
\begin{equation}
E_{\rm tr} \approx {\gamma (Ra_F^*)^{1/2}\over \Gamma^2 k_{\rm mix}D}.
\label{transition}
\end{equation}
At a given $Ra_F^*$, Regime I (or II) occurs for large Ekman
numbers while the asymptotic regime occurs for small Ekman numbers,
with a transition near $E_{\rm tr}$.  Analogously, at constant $E$,
Regime I (or II) dominates at small $Ra_F^*$ while the asymptotic
regime occurs at large $Ra_F^*$, with a transition near modified
flux Rayleigh numbers of $\Gamma^4 \gamma^{-2} (k_{\rm mix} D)^2 E^2 $.

\begin{figure}
\includegraphics[scale=0.4]{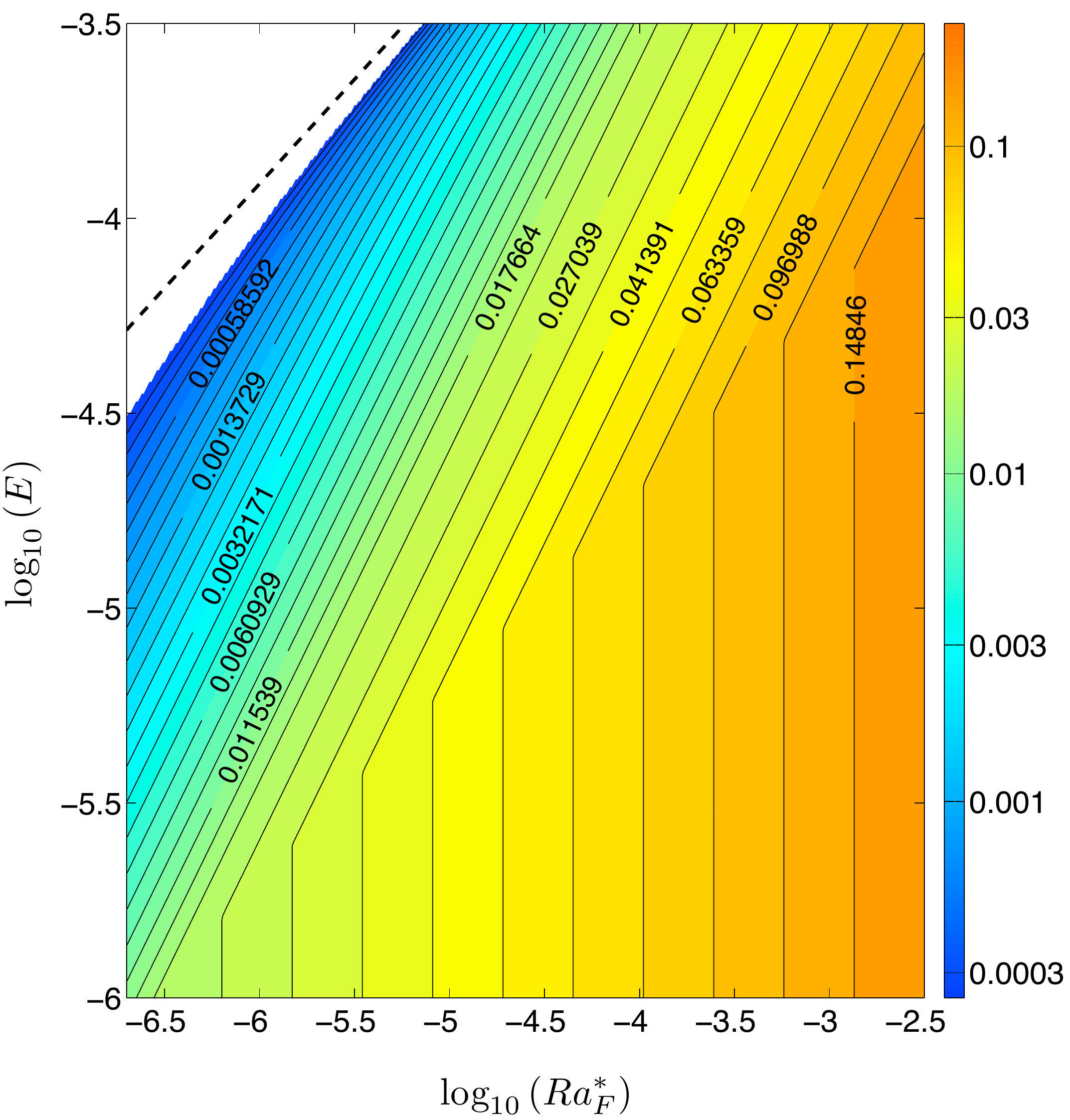}
\caption{Predicted jet speeds for a scaling law that combines
Regimes I, II, and III, given in Eq.~(\ref{three-regimes}),
for the case $\epsilon_{\rm max}=0.3$, $kD=5\pi$, and $\gamma=1$.
The value of $k_{\rm mix}$ is chosen so that the prefactor in
the asymptotic regime, 
$\chi\equiv\epsilon_{\rm max}^{1/2}(k_{\rm mix}D)^{1/2}/(\gamma^{1/2}kD)$,
equals one.
Depicts contours of Rossby number versus modified flux Rayleigh
number (abscissa) and Ekman number (ordinate). 
In the Regimes I and II, contours tilt upward to the right, whereas
in the asymptotic regime they are vertical.  For reference, 
the dashed line depicts the
critical modified flux Rayleigh number given by Eq.~(\ref{Racrit}).
Above the dashed line, convection does not occur.}
\label{three-regimes-2D-plot}
\end{figure}

To illustrate the combined scaling, Fig.~\ref{three-regimes-2D-plot} 
depicts Eq.~(\ref{three-regimes}) as contours of Rossby number 
versus $Ra_F^*$ and $E$ for the specific case $\epsilon_{\rm max}=0.3$,
$\gamma=1$, and $kD=5\pi$ (the same values as in all previous figures).
The value of $k_{\rm mix}$ is chosen so that the prefactor in
the asymptotic regime 
$\epsilon_{\rm max}^{1/2}(k_{\rm mix}D)^{1/2}(\gamma^{1/2}kD)^{-1}
\equiv \chi = 1$.
Regimes II and I lie in the upper left and exhibit sloped contours.
The asymptotic regime lies at the lower right and
exhibits vertical contours, consistent with the expectation
that Rossby number is independent of $E$ there.
As expected from Eq.~(\ref{transition}), the transition between
Regimes I and III slopes gradually upward to the right; the asymptotic
regime extends to larger $E$ when $Ra_F^*$ is larger.
For the value of $k_{\rm mix}$ adopted in 
Fig.~\ref{three-regimes-2D-plot}, Christensen's 
simulations lie predominantly within Regimes I and II; 
however, the predicted transition to the asymptotic regime 
occurs close to Christensen's highest-$Ra_F^*$ cases for each 
value of $E$. If Fig.~\ref{three-regimes-2D-plot} is correct, it 
would suggest that the asymptotic
regime would become clear at Ekman numbers an order of magnitude smaller
than Christensen explored for each value of $Ra_F^*$.  Note, however, that in 
reality the transition between the regimes will probably occur
smoothly across a broad strip straddling $E_{\rm tr}(Ra_F^*)$;
if so, it might be necessary to reach Ekman numbers as low
as $10^{-7}$ to see the asymptotic regime clearly. 

\begin{figure}
\includegraphics[scale=0.4]{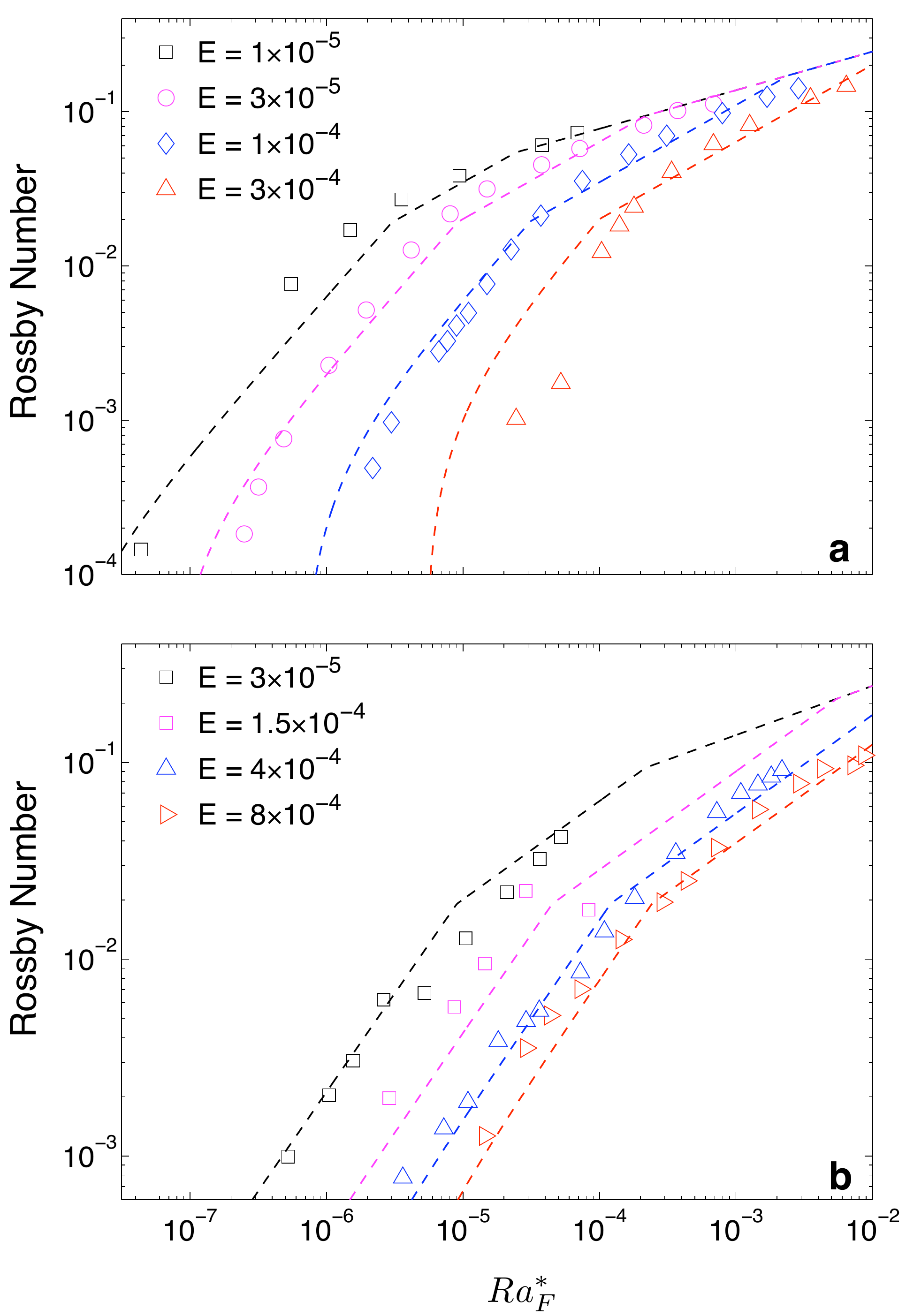}
\caption{Predicted jet speeds for a scaling law that combines
Regimes I, II, and III, given in Eq.~(\ref{three-regimes}),
for the case $\epsilon_{\rm max}=0.3$, $kD=5\pi$, and $\chi=1$.  
Depicts Rossby number versus modified flux Rayleigh number for
several values of $E$ for our combined scaling (curves) and from
simulations (symbols).  Panel {\it a}: \citet{christensen-2002}'s
Boussinesq simulations, at Ekman numbers of $1\times10^{-5}$,
$3\times10^{-5}$, $10^{-4}$, and $3\times10^{-4}$, with scaling
predictions evaluated at those same values (plotted in the corresponding
colors).  Panel {\it b:} Anelastic simulations at Ekman numbers
of $1.5\times10^{-4}$, $4\times10^{-4}$, and $8\times10^{-4}$, along
with a Boussinesq simulation at $3\times10^{-5}$, performed using
the \citet{kaspi-etal-2009} model.  Curves give scaling predictions
evaluated at those same values (plotted in corresponding colors).
The combined scaling matches (to within a factor of $\sim$2)
the mass-weighted mean jet speeds obtained in both the Boussinesq and 
anelastic simulations over a factor of $10^5$ in $Ra_F^*$, $10^2$
in Ekman number, and $10^3$ in Rossby number.}
\label{three-regimes-ro-vs-ra}
\end{figure}

Now we examine Fig.~\ref{three-regimes-ro-vs-ra}{\it a},
which plots our combined scaling (Eq.~\ref{three-regimes}) versus
$Ra_F^*$ for several values of $E$ and compares them to
\citet{christensen-2002}'s Boussinesq simulations.  Curves and
simulation data are shown for $E=10^{-5}$ (black), $3\times10^{-5}$ 
(magenta), $10^{-4}$ (blue), and $3\times10^{-4}$ (red).  The
three regimes are clearly visible in the scalings and agree favorably
with the simulations throughout the plotted range.  
The gradual increase in the slopes in Regime II near the left edge---which 
can be seen in both the scalings and the simulations, especially at 
Ekman numbers of $3\times10^{-5}$ and $10^{-4}$---results 
from the fact that $Ra_F^*$ approaches its critical value (evaluated
for the scalings using Eq.~\ref{Racrit}).
Although the match is not perfect, the scalings not only approximately
reproduce the asymptotic behavior suggested by Christensen but also
exhibit the trend (suggested by Christensen's simulations) where the transition
to the asymptotic regime ocurs at larger $Ra_F^*$ for larger $E$.  The
values of $Ra_F^*$ at these transitions are similar in the scalings
as in Christensen's simulations to within an order of magnitude for most
$E$ values.  Moreover, the actual values of Rossby number predicted by the 
scaling at a given $Ra_F^*$ and $E$ match those of Christensen's simulations 
to within a factor of $\sim$2.

Figure~\ref{three-regimes-ro-vs-ra}{\it b} compares our combined scaling
against simulations performed using the \citet{kaspi-etal-2009} model.
Magenta squares, blue upward-pointing triangles, and red right-pointing
triangles present the Rossby numbers from anelastic simulations
performed respectively
at $E=1.5\times10^{-4}$, $4\times10^{-4}$, and $8\times10^{-4}$
using the Jovian radial structure as in \citet{kaspi-etal-2009},
while the black squares present Rossby numbers from a Boussinesq
simulation at $E=3\times10^{-5}$ performed with the same model.
Scaling predictions are overplotted in the corresponding colors
(black, magenta, blue, and red for Ekman numbers of $3\times10^{-5}$,
$1.5\times10^{-4}$, $4\times10^{-4}$, and $8\times10^{-4}$, respectively).
In these runs, the Prandtl number equals 10, leading to a smaller
critical Rayleigh number than given in Eq.~(\ref{Racrit}); 
as a result, the plotted ranges are
sufficiently supercritical that the scaling predictions do not
exhibit a gradual change in slope near the left edge of the plot 
(in contrast to the behavior in panel {\it a}). Again, the scalings agree
well with the simulations; the discrepancy is everywhere less
than a factor of 2.  This agreement is encouraging considering that,
together, Fig.~\ref{three-regimes-ro-vs-ra}{\it a} and {\it b}
present results spanning factors of 200,000 in $Ra_F^*$, 
1000 in $Ro$, and 80 in $E$.

Importantly, when the modified flux Rayleigh number is defined
in the appropriate mass-weighted way (Eq.~\ref{Ra_Fstar-integrated}), 
the {\it same scalings} explain the mass-weighted, global-mean Rossby numbers
of {\it both} the Boussinesq and anelastic simulations, despite the
fact that the former use constant mean density while the latter
adopt background densities (and thermal expansivities) that
vary radially by several orders of magnitude from top to bottom.
The Boussinesq simulations exhibit relatively little axial shear of
the zonal wind along columns of constant $s$ \citep{christensen-2002},
while the anelastic simulations described here exhibit significant
shear \citep{kaspi-etal-2009}.  The fact that our scaling arguments
can explain both sets of runs indicates that this shear (if present)
does not control the mass-weighted mean interior wind speed.

Interestingly, Fig.~\ref{three-regimes-2D-plot} suggests that
Regime I disappears for sufficiently small Ekman and modified
flux Rayleigh numbers ($E\le 10^{-7}$
and $Ra_F^* \le 10^{-8}$ for the parameters chosen).  If so, 
then at very small Ekman numbers, Regime II abuts directly against 
Regime III.  No numerical simulations have yet been performed at these
small parameter values, however, and 
how this transition occurs remains an open question. Equation~(\ref{three-regimes}) may need modification in 
this low-$Ra_F^*$ and low-$E$ portion  of the parameter space 
pending further numerical and theoretical work.

Next, we extrapolate the scalings to Jupiter, where the
Ekman and modified flux Rayleigh numbers are
$\sim$$10^{-15}$ and $\sim$$10^{-13}$--$10^{-14}$, respectively.
Despite the uncertainties, 
Eq.~(\ref{three-regimes}) would suggest that Jupiter lies in the
asymptotic regime.  For the Jovian values of $Ra_F^*$ and $E$, 
the scalings predict Rossby numbers of 
$\sim$$3\times10^{-5}$--$6\times10^{-4}$ and the implied zonal 
wind speeds in the interior are $\sim$0.1--$1\rm\,m\,sec^{-1}$.  
Although numerical simulations that are overforced by orders of
magnitude can easily produce Jupiter-like jet speeds \citep{christensen-2001,
christensen-2002, heimpel-aurnou-2007}, both Christensen's
asymptotic fit and our mixing-length scalings therefore
suggest that, given the weakly forced conditions of Jupiter's
interior, the zonal-jet speeds in the interior are much slower
than the observed cloud-level values (which reach $\sim$$150\rm\,m\,sec^{-1}$
for Jupiter and $400\rm\,m\,sec^{-1}$ for Neptune).  {\tt Nevertheless,
this conclusion is tentative, because our estimates of eddy viscosity
are uncertain and because the asymptotic regime (if it exists!)
is not yet obvious in numerical simulations.  Stronger confirmation of
the existence or absence of the asymptotic regime in numerical models is 
necessary.}

\section{Conclusions}
\label{conclusions}

Over the past two decades, several authors have performed three-dimensional
numerical simulations of low-viscosity convection in rapidly rotating
spherical shells to test the hypothesis
that convection in the molecular interiors pumps the fast zonal jet
streams observed on Jupiter, Saturn, Uranus, and Neptune.
These studies show that the zonal jet speeds can range over 
many orders of magnitude depending on the heat flux,
viscosity, and other parameters \citep[e.g.,][]{christensen-2002,
kaspi-etal-2009}. Here, we presented 
simple theoretical arguments to explain the characteristic jet 
speeds---and their dependence on heat flux and viscosity---seen
in these studies.

Our main findings are as follows:
\begin{itemize}

\item We demonstrated that the characteristic convective
velocities in the \citet{kaspi-etal-2009} simulations
scale well with $(\alpha g F/\rho c_p \Omega)^{1/2}$.
While such a rotating scaling has long been proposed 
and is consistent with rotating laboratory experiments 
\citep[e.g.,][]{fernando-etal-1991}, this to our knowledge is
the first demonstration of its applicability to the interiors
of giant planets over a wide range of heat fluxes and rotation
rates.  

\item
Next, we attempted to explain the jet speeds obtained by
\citet{christensen-2002} and \citet{kaspi-etal-2009} in
three-dimensional numerical simulations of convection with free-slip
boundaries.  At weakly supercritical Rayleigh numbers,
linear theory \citep[e.g.,][]{busse-hood-1982, busse-1983} and 
numerical simulations \citep[e.g.,][]{christensen-2002} show
that the eastward and outward convective velocity components 
are highly correlated outside the tangent cylinder.  
We showed that if the degree of correlation $C$
between the eastward and outward convective velocity components 
is constant (as expected at very low supercriticalities)
and if the jets are damped by a numerical viscosity, then
the mean jet speeds should scale as the convective heat flux over
the viscosity (Regime II).  This scaling explains the jet speeds obtained in
both Boussinesq and anelastic simulations at low Rayleigh
numbers, where the jet speeds are relatively weak.

\item
On the other hand, at higher Rayleigh number, the eastward
and outward convective velocity components become increasingly
decorrelated.  We showed that, 
if the fraction of convective energy release used to pump the jets 
($\epsilon$) is a constant in this regime, and if the jets are damped by a 
numerical viscosity, then the mean jet speeds should scale as 
$(F/\nu)^{1/2}$.    This scaling (Regime I) explains quite well the
mean jet speeds obtained in the simulations from \citet{christensen-2002}
and \citet{kaspi-etal-2009} in cases when the jets dominate the total
kinetic energy, as occurs at highly supercritical Rayleigh numbers.

\item
The relationship between the correlation coefficient $C$ and the
jet-pumping efficiency $\epsilon$ shows how the transition between
these two regimes can naturally occur.  A nearly constant correlation
coefficient implies that the jet-pumping efficiency must increase
strongly with $Ra_F^*$.  When the jet-pumping efficiency finally
plateaus near its maximum value of 1, then the correlation 
coefficient must begin to decrease strongly with
further increases in $Ra_F^*$ (see Eq.~\ref{epsilon3}).  Thus,
a natural regime transition occurs between Regime II at low
Rayleigh numbers and Regime I at high Rayleigh numbers.  Moreover,
we rigorously calculated $C$ and $\epsilon$ from the anelastic
simulations and showed that, as expected, $C$ decreases and 
$\epsilon$ increases with $Ra_F^*$.  A simple,
heuristic theory that transitions smoothly between constant
correlation coefficient at low $Ra_F^*$ and a constant jet-pumping
efficiency at high $Ra_F^*$ explains the qualitative features
of the correlation coefficient and jet-pumping efficiency as
calculated from the simulations.

\item Both the Boussinesq \citep{christensen-2002} and
anelastic simulations hint at the existence of a third regime
where, at sufficiently large heat flux and/or sufficiently
small viscosity, the characteristic jet speeds become independent
of the viscosity.  We constructed a simple mixing-length scaling 
for the behavior in this regime under the 
assumption that the jet-pumping efficiency is constant
and the damping results from an eddy viscosity (rather
than the molecular or numerical viscosity).  This scaling
suggests that the mean jet speed in the
molecular interior should scale with the convective heat flux to the 1/4 
power.  Given the simplicity of our model, this agrees favorably 
with the asymptotic fit suggested by \citet{christensen-2002}, 
which proposes that jet speeds should scale with the convective
heat flux to the 1/5 power.  Both scalings suggest---{\tt tentatively}---that the mean
wind speeds in the molecular interior should be significantly
weaker than the jet speeds measured at the cloud level. 

\item A scaling that combines these three regimes
can explain (to within a factor of $\sim$2) the mass-weighted 
mean jet speeds obtained in both the Boussinesq and anelastic 
simulations ranging across a factor of $10^5$ in $Ra_F^*$, $10^2$ in Ekman
number, and $10^3$ in Rossby number.
Importantly, when the modified flux Rayleigh number is defined in
an appropriate mass-weighted way, the {\it same scalings} apply to
both Boussinesq and anelastic cases.


\end{itemize}

It is important to emphasize that the scaling theories presented
in Sections \ref{viscous}--\ref{combine} apply only to the
mass-weighted mean wind speeds.  The theory
does not provide information on the detailed three-dimensional
structure of the jets, such as their variation in latitude.

Several challenges remain for the future.  First, we did
not consider the effect of the Prandtl number on the jet speeds,
mostly because a careful characterization of the $Ra_F^*$- and
$E$-dependence of the jet speeds over a wide range has only
been performed for a very limited number of $Pr$ values 
(1 for \citet{christensen-2002} and 10 for \citet{kaspi-etal-2009}).  
However, several studies show that $Pr$ does affect the jet speeds 
\citep[e.g.,][]{aubert-etal-2001, christensen-2002, kaspi-2008},
although one would expect this dependence to disappear in the
asymptotic limit.   Extending our scalings to include the $Pr$ 
dependence is an important goal for future work.  

Second, the simulations we set out to explain---those of 
\citet{christensen-2002} and \citet{kaspi-etal-2009}---assume
the convection occurs in a thick shell, for which the region 
outside the tangent cylinder dominates the total mass, volume, 
and kinetic energy.   As shown by several 
authors, however, the dynamical mechanisms and details of jet pumping 
differ for the fluid inside and outside the tangent cylinder
\citep[e.g.,][]{heimpel-aurnou-2007}.  Thus, the scaling properties 
could differ for simulations in a thin shell \citep{heimpel-etal-2005, 
heimpel-aurnou-2007}, for which the high-latitude regions inside the 
tangent cylinder will more strongly influence mass-weighted mean flow
properties.

Third, we assumed that the wavenumber $k$---which represents the length 
scale for the variation of zonal-wind shear and Reynolds stress with 
cylindrical radius $s$---was a constant with the Rayleigh and
Ekman numbers.  To zeroth order, this seems a reasonable
approximation, because simulations suggest that 
the widths of the jets outside the tangent cylinder are set by 
the adopted shell thickness and not by the Ekman and Rayleigh 
numbers.\footnote{In the equatorial plane, the simulations generally
exhibit westward flow at the inner boundary and eastward flow at the 
outer boundary, with a relatively smooth transition in between.}
Nevertheless, analytic solutions in the linear limit suggest that the 
{\it zonal} wavelengths near the critical Rayleigh number should
scale as $E^{1/3}$ \citep[e.g.,][]{roberts-1968, busse-1970},
and scaling arguments in the context of convection with no-slip
boundaries suggest that the sizes of vortex rolls scale as $E^{1/5}$ 
\citep{aubert-etal-2001}.  Thus, the wavenumber representing
the $s$-variation of the zonal-jet-shear and Reynolds stresses
could potentially have a weak dependence on $E$.
Although our assumption of constant $k$
seems to work reasonably well over the simulated range, it would
be worth relaxing this assumption in the future.

Our scaling arguments---and the simulations they attempt to 
explain---assume the fluid is electrically neutral.  However,
the interiors of Jupiter and Saturn become metallic at pressures
exceeding $\sim$$1\,$Mbar \citep{guillot-etal-2004}, and the
resulting magnetohydrodynamic effects could influence the jet
dynamics even in the molecular region \citep[e.g.,][]{kirk-stevenson-1987,
liu-etal-2008, glatzmaier-2008}.  For example, \citet{liu-etal-2008} 
argued that Ohmic dissipation
would limit strong jets to regions above 96\% (86\%) of the
radius on Jupiter (Saturn).  They further argued that, if
the interior is isentropic and exhibits columnar Taylor-Proudman
behavior, the winds would be weak not only in the electrically
conducting region but throughout the convection
zone.  This result is consistent with ours although
it invokes different physics.  Moreover, the simulations we investigated
do not include solar forcing or latent heating.  These effects may
induce significant internal density perturbations---and therefore
thermal-wind shear---within a few scale heights of the observable
cloud deck \citep[e.g.,][]{williams-2003a, kaspi-flierl-2007, 
lian-showman-2008, lian-showman-2010, schneider-liu-2009}.
The development of convective models that include these effects is a major
goal for the future.


%




\begin{acknowledgments}
This work was supported by NSF grant AST-0708698 and NASA grants
NNX07AF35G and NNX10AB91G to APS, NSF grant AST-0708106 to GRF, and a NOAA Climate
and Global Change Postdoctoral Fellowship to YK administered by the 
University Corporation for Atmospheric Research.
\end{acknowledgments}

\def\planss{Planet. Space Sci.}
\def\pre{Phys. Rev. E}
\bibliography{showman-bib}

\end{article}
\end{document}